\newcommand{\be}{\begin{equation}}
\newcommand{\ee}{\end{equation}}
\newcommand{\bea}{\begin{eqnarray}}
\newcommand{\eea}{\end{eqnarray}}

\documentclass[a4paper,11pt,onecolumn]{article}
\pdfoutput=1
\usepackage{jheppub}
\usepackage[T1]{fontenc}
\usepackage{mlmodern}
\usepackage{graphicx}
\usepackage{dcolumn}
\usepackage{bm}
\usepackage{amsmath}
\usepackage{amssymb}
\usepackage{slashed}
\usepackage{accents}
\def\d{d\kern-.8 ex\vrule height 1.3 ex depth-1.24 ex width .7 ex \kern .15 ex}
\def\D{D\kern-1.7 ex\vrule height .87 ex depth-.8 ex width .7 ex \kern .95 ex}

\title{Diffusion and instabilities in large-N holographic Fermi liquids: the vector fluctuations of the electron star}
\author{Vladan Gecin}
\author{and Mihailo \v{C}ubrovi\'c}
\affiliation{Center for the Study of Complex Systems, Institute of Physics Belgrade, University of Belgrade, Pregrevica 118, 11080 Belgrade, Serbia}
\emailAdd{gecin@ipb.ac.rs}
\emailAdd{cubrovic@ipb.ac.rs}

\date{\today}

\abstract{
We study the hydrodynamic response of the AdS electron star in the vector sector, and compute the correlation functions and the transverse conductivity of the dual field theory. The system exhibits hydrodynamic behavior at low temperatures and near the critical temperature where the electron star undergoes the phase transition to the RN black hole. However, at intermediate temperatures the hydrodynamics does not exist. Remarkably, the system has an instability, i.e. a pole on the positive imaginary frequency axis at finite temperature.  This instability is found both from analytical arguments and from numerics. Its physical meaning is so far unclear but it might mean that the ideal fluid limit for the star is a false vacuum.}

\begin{document}

\maketitle

\section{Introduction}\label{secintro}

Holographic studies of strongly correlated electron systems are presently a somewhat forgotten field, although at some point they were among the most popular directions in AdS/CFT \cite{Vegh:2009,Leiden:2009,Faulkner:2010zz}. Although holography has given rise to a number of important ideas, such as semilocal quantum liquids \cite{Liu:2011,Gubser:2009qt}, fractionalized vs.~coherent fermionic matter \cite{Huijse:2011hp} and the holographic lattices \cite{Smit:2021dwh,Hercek:2022wyu,Balm:2022bju}, our impression is that surprisingly little has been revealed on the fundamental issue of how exactly robust Fermi surfaces and robust fermionic phases (be it Fermi liquids or strange metals) arise from the basic properties of the system such as symmetries and thermodynamics. This is supposed to be the strong point of holography: without describing the microscopics of finite-density systems (QCD, heavy ions, superfluids, (non-)Fermi liquids...), it provides an effective field-theory description in the form of a ``scaling atlas'' \cite{Charmousis:2010zz,Gouteraux:2011ce,Gouteraux:2012yr,Zaanen:2021llz}. It is clear that the field needs a fresh start, from the viewpoint of an effective field theory which is able to directly capture the key phenomenology of Fermi surfaces. But before we are ready for such an endevaour, we want to fully understand the extant holographic Fermi surface models, and pinpoint what exactly they are missing.

More specifically, we want to understand the physics of the simplest holographic Fermi liquid model -- the electron star -- by studying its hydrodynamic response. For all we know, electron stars represent something like multiplets of flavored Lifshitz Fermi liquids. We may indeed call them Fermi liquids as they have a sharp Fermi surface and long-living quasiparticles, but they are certainly not Landau Fermi liquids -- the thermodynamic quantities and the self-energy of the quasiparticle scale very differently, and there is a large number (strictly speaking infinity) of Fermi surfaces. So far, works in this direction include the brief consideration of conductivity in the original paper \cite{Hartnoll:2010gu} and the study of quasinormal modes and hydrodynamics in \cite{Kaplis:2013gux,Gran:2018jnt}. In \cite{Gran:2018jnt}, the authors consider the hydrodynamic response in the longitudinal sector but the focus is on modes with Robin boundary conditions encoding for a double-trace deformation corresponding to a plasmon \cite{Gran:2017jht,Gran:2018vdn}. Nevertheless, even with standard (Dirichlet) boundary conditions, \cite{Gran:2018jnt} finds some modes with exotic dispersion relations (non-diffusive scaling, modes hitting the real axis at specific momenta, etc), a hint that some unusual phenomena might be expected. In \cite{Kaplis:2013gux}, the author studies the diffusive quasinormal mode and finds an anomalous temperature dependence of the diffusion coefficient.

On the other hand, the linear response of AdS${}_2$ metals, where the fermionic response is dominated by the universal near-horizon AdS${}_2$ region of the Reissner-Nordstrom (RN) black hole, is very well studied \cite{Edalati:2010hk,Davison:2011uk,Davison:2013bxa,Davison:2013uha} and is known to be hydrodynamic even at zero temperature, likely because of the excess of massless modes due to the near-horizon $SL(2,\mathbb{R})$ symmetry. The Lifshitz geometry is likewise known to exhibit hydrodynamic response \cite{Sybesma:2015oha,Gursoy:2016tgf}. Electron stars at finite temperature (electron clouds) essentially combine these two sectors -- is there anything new in their response compared to Lifshitz and RN?

Inspired by these works, we systematically compute the response functions and dispersion relations of the transverse sector (as we mention above, the longitudinal sector was studied in \cite{Kaplis:2013gux,Gran:2018jnt} and will be studied further in a separate work) for various fermion charges and various temperatures, including $T=0$. Our intention is to learn about the true nature of this somewhat pathological large-$N$ system using the hydrodynamic response as a diagnostic tool. While we do not think that the electron star as such is a very realistic model of any condensed matter system, it is important to understand its workings in order to construct a more meaningful model. In hindsight, we will find some big surprises, in particular an instability at intermediate temperatures, suggesting that the simple fluid approximation is a false vacuum in some parameter range. 

At this place it is useful to fix some terminology. The zero-temperature solution is universally called the electron star. The finite-temperature variant was originally called simply the electron star at finite temperature \cite{Hartnoll:2010ik} but later on, e.g. in \cite{GiangrecoMPuletti:2015grq,Gran:2018jnt}, it was called the electron cloud, a more illustrative term as it captures the fact that at finite temperature the electron density is only nonzero at some distance from the horizon. We thus exclusively use the term electron cloud for the thermal case and reserve the name electron star solely for the zero-temperature case.

The structure of the paper is the following. In Section \ref{secback} we briefly recapitulate the story of electron stars and electron clouds from \cite{Hartnoll:2010gu,Hartnoll:2010ik} and describe their numerical construction. In Section \ref{secfluc} we construct the linearized fluctuation equations in the transverse sector, define the boundary conditions and compute the response functions. Section \ref{secresults} brings the main physical results: the structure of the response functions, the presence of the hydrodynamic response and the instability. These numerical findings are largely analytically reproduced in Section \ref{secanal}. Section \ref{secconc} summarizes the conclusions.

\section{Electron star backgrounds}\label{secback}

\subsection{The fluid limit and the equations of motion}

It is known that a charged AdS black hole in the presence of a fermionic field becomes unstable \cite{Hartnoll:2009ns}: it explodes into a charged star, which is thermodynamically preferred to the charged black hole because the electric repulsion of a less compact object is less energetically costly. The approach taken in \cite{Hartnoll:2010gu} that makes calculations easy is to consider the Thomas-Fermi limit for the fermions: they become a semiclassical fluid and their stress-energy tensor is easy to write, requiring no loop calculations. The quantumness of the fluid is present only through its equation of state, otherwise it is a classical object with sharp boundaries, called electron star at zero temperature or electron cloud at finite temperature.

To realize the above idea, one starts from the microscopic Einstein-Maxwell-Dirac action in asymptotically AdS${}_4$ spacetime of radius $L$:
\begin{equation}
S=S_\mathrm{g}+S_\mathrm{EM}+S_\mathrm{Dirac},\label{2s}
\end{equation}
where the Einstein, Maxwell and Dirac component are given respectively by
\begin{eqnarray}
S_\mathrm{g}&=&\frac{1}{2\kappa^2}\int \mathrm{d}^4x\sqrt{-g}\left(R+\frac{6}{L^2} \right),\label{2sg}\\
S_\mathrm{EM}&=&-\frac{1}{4e^2}\int \mathrm{d}^4x\sqrt{-g} F_{\mu\nu}F^{\mu\nu},\label{2sem}\\
S_\mathrm{Dirac}&=&-\frac{L^2}{\kappa^2}\int\mathrm{d}^4x\sqrt{-g}\bar{\Psi}\left(\Gamma^\mu D_\mu-mL\right)\Psi.\label{2sdirac}
\end{eqnarray}
Here, $\Psi$ is the Dirac spinor with charge $e$ and mass $m$, the covariant derivative includes the spin connection as $D_\mu=\partial_\mu+\omega_{\mu A B}\Gamma^{AB}/4-i (e L/\kappa) A_\mu$, and $\kappa$ is the gravitational coupling constant. Taking the Tolman-Oppenheimer-Volkov fluid approximation yields the electron star, the charged AdS analogue of the familiar neutron star, which is appropriate when the number of occupied levels is very large. In this limit we can replace the microscopic Dirac action by the action of a fluid with energy density $\rho$, charge density $\sigma$ and pressure $p$, given in \cite{Hartnoll:2010gu}:
\be
S_\mathrm{fluid}=\int \mathrm{d}^4x\sqrt{-g}\left(-\rho\left(\sigma\right)+\sigma u^{\mu}\left(\partial_{\mu}\phi+A_{\mu}\right)+\lambda\left(u^2+1\right)\right).\label{2sfluid}
\ee
Here, $\phi$ is the auxiliary (Clebsch) potential and $\lambda$ is a Lagrange multiplier (for details see \cite{Hartnoll:2010gu}; we will not make explicit use of this action). In the ideal fluid approximation, the above general fluid action further simplifies to just the pressure $p$ of the fluid when calculated on-shell. The summation over the Fermi sea (i.e., over the occupied levels of the fermions) now becomes the integration, yielding closed-form expressions for densities and pressure:\footnote{In this approximation the results for $\rho$, $\sigma$ and $p$ are not sensitive to curvature and have the same form as in flat space \cite{Hartnoll:2010gu}.}
\begin{equation}
\hat{\rho}=\hat{\beta}\int_{\hat{m}}^{\hat{\mu}_\mathrm{loc}}\epsilon^2\sqrt{\epsilon^2-\hat{m}^2}\mathrm{d}\epsilon,\quad \hat{\sigma}=\hat{\beta}\int_{\hat{m}}^{\hat{\mu}_\mathrm{loc}}\epsilon\sqrt{\epsilon^2-\hat{m}^2}\mathrm{d}\epsilon,\quad -\hat{p}=\hat{\rho}-\hat{\mu}_\mathrm{loc}\hat{\sigma},\label{2rhosigmap}
\end{equation}
where the hats denote dimensionless quantities, defined as
\begin{equation}
    \hat{p}=L^2\kappa^2p,\quad \hat{\rho}=L^2\kappa^2\rho,\quad \hat{\sigma}=e L^2\kappa\sigma,\quad \hat{\beta}=\frac{e^4L^2}{\kappa^2}\beta,\quad \hat{m}^2=\frac{\kappa^2}{e^2}m^2,\quad \hat{\mu}_\mathrm{loc}=\frac{\kappa}{e}\mu_\mathrm{loc},\label{2hats}
\end{equation}
and where we define the local \emph{bulk} chemical potential as \cite{Hartnoll:2010gu}
\begin{equation}
\mu_\mathrm{loc}=\frac{A_t}{\sqrt{-g_{tt}}}\label{2muloc}.
\end{equation}
We thus have two microscopic free parameters to vary: the fermion mass $\hat{m}$ and the level density constant of the fermion $\hat{\beta}$. The fermion mass must satisfy the inequality $0\le \hat{m}<1$ \cite{Hartnoll:2010gu}, while $\hat{\beta}$ is in principle arbitrary, but cannot be too large if the fluid approximation is to be valid \cite{Leiden:2011}. Taking into account the requirement of string theory that the gravitational coupling equals the square of the Maxwell coupling, we may assume that in the classical gravity regime:
\begin{equation}
    e^2\sim\frac{\kappa}{L}\ll 1,
\end{equation}
which means that $\hat{\beta}\sim 1$ \cite{Hartnoll:2010gu}.

One last ingredient we need before solving the equations of motion is the counterterm to the action (\ref{2s}): $S\mapsto S+S_\mathrm{ct}$. The counterterm ensures that we have a good action principle and eliminates divergences. It reads \cite{Emparan:1999pm}:
\begin{equation}
S_\mathrm{ct}=-\frac{1}{\kappa^2}\int\mathrm{d}^3x\sqrt{-\gamma}\left(K+\frac{2}{L}+\frac{L}{2}R^{(3)}\right).
\end{equation} 
Here $\gamma$, $R^{(3)}$ and $K$ are the induced metric and the Ricci scalar on the boundary and the trace of extrinsic curvature, respectively. While not necessary for the on-shell (background) solution, the counterterm is crucial for the computation of two-point correlators from fluctuation equations.

We can now adopt an ansatz for the solution to construct the equations of motion. Assume the following form for the metric and gauge field:
\begin{equation}
    ds^2=L^2\left(-f(r)dt^2+g(r)dr^2+\frac{1}{r^2}(dx^2+dy^2)\right),\quad A=\frac{e L}{\kappa}h(r)\mathrm{d}t.
\end{equation}
This ansatz encapsulates the most general radially symmetric static solution, homogeneous and isotropic along the transverse coordinates $(x,y)$, with no magnetic field or stationary current. The AdS boundary is located at $r=0$. Now, finally, from Eqs.~(\ref{2s}), (\ref{2sfluid}) and (\ref{2rhosigmap}-\ref{2muloc}) and the above ansatz, the equations of motion are found to be \cite{Hartnoll:2010gu}:
\begin{eqnarray}
\frac{1}{r}\left( \frac{f'}{f}+\frac{g'}{g}+\frac{4}{r} \right)+\frac{g h \hat{\sigma}}{\sqrt{f}}&=&0,\label{2eom1}\\
\frac{f'}{r f}-\frac{{h'}^2}{2f}+g(3+\hat{p})-\frac{1}{r^2}&=&0,\label{2eom2}\\
h''+\frac{g\hat{\sigma}}{\sqrt{f}}\left( \frac{r h h'}{2}-f \right)&=&0.\label{2eom3}
\end{eqnarray}
We will now discuss their solutions at zero temperature and at finite temperature.

\subsection{Solutions}
\subsubsection{Zero-temperature solutions: electron stars}

At zero temperature, the only infrared (IR) boundary condition is the smoothness of the solution in the interior ($r\to\infty$). The low-energy (large-$r$) behavior of such solution is well-known: it is a perturbed Lifshitz geometry with the scaling exponent $z$ ($1\leq z<\infty$) \cite{Hartnoll:2010gu}. It is obtained by solving the equations of motion (\ref{2eom1}-\ref{2eom3}) order by order in the large-$r$ limit and reads:
\begin{equation}
    f_{\mathrm{in}}=\frac{1}{r^{2z}}(1+f_1 r^{\alpha}+\ldots),\quad g_{\mathrm{in}}=\frac{g_{\infty}}{r^2}(1+g_1 r^{\alpha}+\ldots),\quad h_{\mathrm{in}}=\frac{h_{\infty}}{r^z}(1+h_1 r^{\alpha}+\ldots).
\end{equation}
The constants $g_{\infty}$, $h_{\infty}$, $g_1$, $h_1$ and the power $\alpha$ are obtained by solving an algebraic set of equations and depend on $z$, $\hat{m}$ and $\hat{\beta}$ (which is an implicit function of $z$ and $\hat{m}$ itself), while $f_1$ has to be negative and can be otherwise set in an arbitrary way. Following \cite{Hartnoll:2010gu}, we set $f_1\equiv-1$, at the expense of rescaling the coordinates. This geometry is to be matched numerically with that of the outer sector. The edge of the star, where the matching is to be done, is at the radius $r_s$ where the local chemical potential cannot accommodate even a single fermion anymore. Its radius is thus determined by the relation
\begin{equation}
\frac{h(r_s)}{\sqrt{f(r_s)}}=\hat{m}.\label{2rs}
\end{equation}
Having zero matter density, the outer sector is naturally described by the RN geometry:
\begin{equation}
f_\mathrm{out}=c^2\left( \frac{1}{r^2}-\hat{M}r+\frac{1}{2}\hat{Q}^2r^2 \right),\quad g_\mathrm{out}=\frac{c^2}{r^4f_\mathrm{out}(r)},\quad h_\mathrm{out}=c(\hat{\mu}-\hat{Q}r).\label{2rnouter}
\end{equation}
The speed of light $c$ had to be included because of the rescaled time coordinate, while the dimensionless constants $\hat{M}$, $\hat{Q}$ and $\hat{\mu}$ are the total mass and the total charge of the star and the field theory chemical potential, respectively.\footnote{Note that for the outer RN sector we use the conventions of \cite{Hartnoll:2010ik}.}  The star solution is given in Figures \ref{fig:essol} and \ref{fig:essigma} (left panels).

\subsubsection{Finite-temperature solutions: electron clouds}

At finite temperature the star is not a ball anymore, but rather a spherical shell surrounding a black hole. It is appropriately called ``electron cloud'' in \cite{Gran:2017jht,Gran:2018jnt}, so we have adopted this terminology.
Its interior is described by the RN solution \cite{Hartnoll:2010ik}:
\begin{equation}
f_\mathrm{in}=\frac{1}{r^2}-\left( \frac{1}{r^2_+}+\frac{\hat{\mu}^2_0}{2} \right)\frac{r}{r_+}+\frac{\hat{\mu}^2_0}{2}\left(\frac{r}{r_+}\right)^2,\quad g_\mathrm{in}=\frac{1}{r^4f_\mathrm{in}(r)},\quad h_\mathrm{in}=\hat{\mu}_0\left(1-\frac{r}{r_+} \right),\label{2rninner}
\end{equation}
where $\hat{\mu}_0$ is a constant unrelated to the chemical potential in field theory, while $r_+$ is the outer event horizon of the RN black hole. Without loss of generality we can set $r_+=1$. The inner boundary of the star (cloud) is located at a radius $r_1$ where the local chemical potential equals the fermion mass (this is when we start filling the Fermi sea, as one can see from the expressions (\ref{2rhosigmap})). Likewise, the outer boundary of the star is located at the radius $r_2$ where the local chemical potential can no longer accommodate fermions (being determined in the same way as $r_s$ at $T=0$), and the outer sector ($r<r_2$) is again described by the RN geometry (\ref{2rnouter}). The dimensionless constants $\hat{M}$ and $\hat{Q}$ now correspond to the total mass and the total charge, respectively, of the partially fractionalized system \emph{black hole + fermion fluid}. 

When everything is said and done, we have the solution as
\begin{equation}
    (f,g,h)=\left\{ \begin{array}{ll}
         \left(f_\mathrm{in},g_\mathrm{in},h_\mathrm{in}\right),& 1>r\ge r_1\\
         \left(f_\mathrm{ES},g_\mathrm{ES},h_\mathrm{ES}\right),& r_1\ge r\ge r_2\\
         \left(f_\mathrm{out},g_\mathrm{out},h_\mathrm{out}\right),& r_2\ge r>0
    \end{array} \right.
\end{equation}
The two RN solutions (inner and outer) are given by the analytical expressions (\ref{2rninner}) and (\ref{2rnouter}), while the star solution $(f_\mathrm{ES},g_\mathrm{ES},h_\mathrm{ES})$ has to be determined numerically and sewed together with the inner and the outer region.

The electron cloud exists for $0<T<T_c$, where $T_c$ is some critical temperature. At $T_c$ the cloud vanishes and only the RN black hole remains. This means that for $T>T_c$ all the charge is distributed over the horizon. The holographic dual of this phase is interpreted as the fully fractionalized semilocal quantum liquid, conjectured to describe a strange metal \cite{Vegh:2009,Faulkner:2009wj,Liu:2011}.

\begin{figure}[h]
    \centering
    \begin{minipage}[b]{0.5\textwidth}
    \centering
        \includegraphics[height=0.2\textheight]{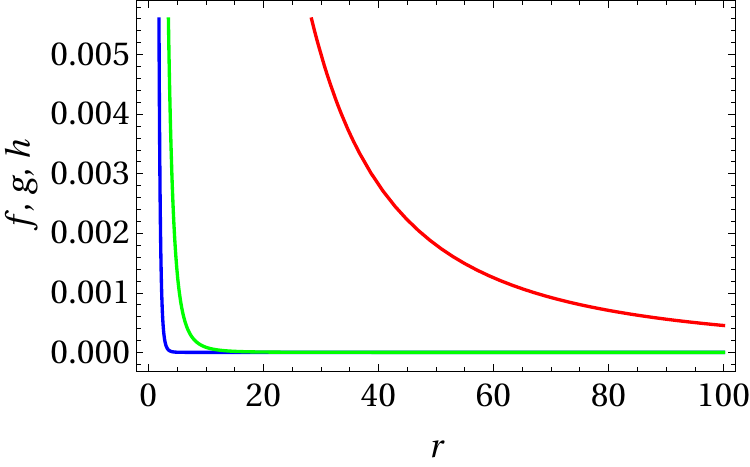}
    \end{minipage}%
    \begin{minipage}[b]{0.5\textwidth}
    \centering
        \includegraphics[height=0.2\textheight]{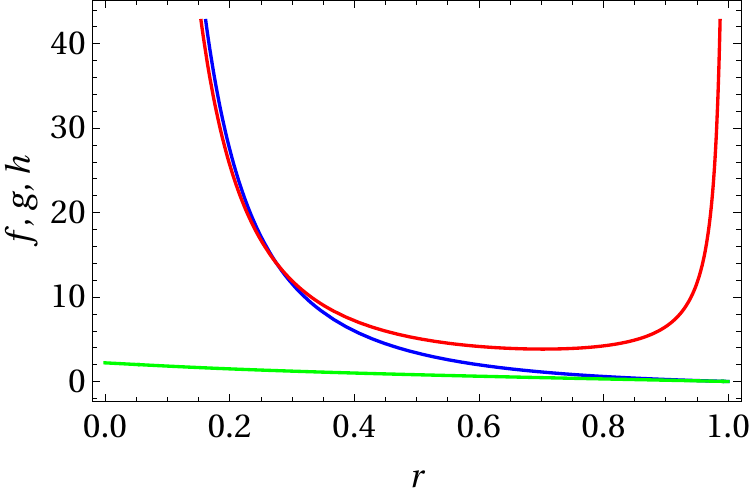}
    \end{minipage}
    \caption{The background functions $f$, $g$ and $h$ (blue, red, green) for $z=4$, for $T=0$ (left) and $T/T_c=0.65$ (right).}
    \label{fig:essol}
\end{figure}

\begin{figure}[h]
    \centering
    \begin{minipage}[b]{0.5\textwidth}
    \centering
        \includegraphics[height=0.2\textheight]{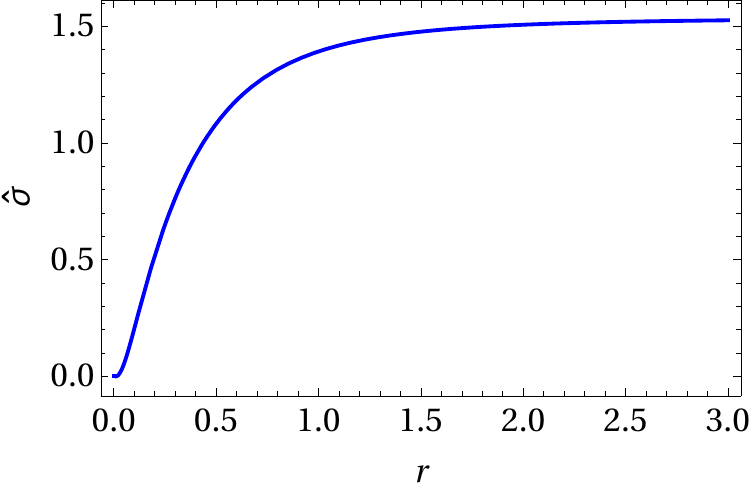}
    \end{minipage}%
    \begin{minipage}[b]{0.5\textwidth}
    \centering
        \includegraphics[height=0.206\textheight]{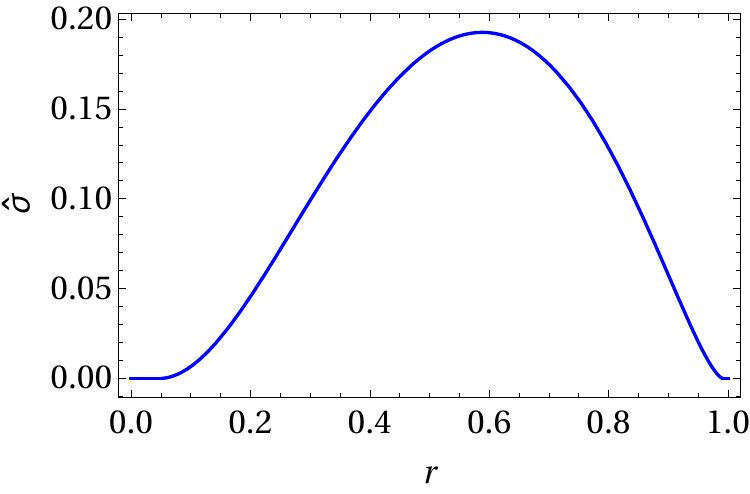}
    \end{minipage}
    \caption{The dimensionless star density $\hat{\sigma}$ for $z=4$, for $T=0$ (left) and $T/T_c=0.65$ (right).}
    \label{fig:essigma}
\end{figure}

At the critical temperature we have $r_1=r_2\equiv r_c$. It turns out \cite{Hartnoll:2010ik} that the critical radius and the critical temperature are given by the condition that the cloud contains just a single fermion, located at the maximum of the bulk chemical potential. This yields the equations for $r_c$ and $T_c$:
\begin{equation}
\frac{h_\mathrm{in}(\hat{\mu}_0(T_c),r_c)}{\sqrt{f_\mathrm{in}(\hat{\mu}_0(T_c),r_c)}}=\hat{m},\quad \frac{\mathrm{d}}{\mathrm{d}r}\frac{h_\mathrm{in}(\hat{\mu}_0(T_c),r_c)}{\sqrt{f_\mathrm{in}(\hat{\mu}_0(T_c),r_c)}}=0,\label{2rc}
\end{equation}
where it is actually understood that one first finds the critical values of $\hat{\mu}_0$ and $r$ satisfying the above equations, and then the critical temperature $T_c$, using the relation
\begin{equation}
    T=\frac{|f'(r_+)|}{4\pi c}.
\end{equation}
The speed of light is included again because of the normalization of the IR solution, i.e. the inner solution (\ref{2rninner}). Since we do not know its value in advance, when we calculate the background functions numerically, we do it by setting the ratio $T/T_c$. That is why we need to take the relations (\ref{2rc}) into account: by setting $T/T_c$, we determine (the non-critical) $\hat{\mu}_0$, whence we determine the inner radius $r_1$ of the star. An example of the electron cloud solution is given in the right panel of the Figures \ref{fig:essol} and \ref{fig:essigma}.


\section{Fluctuation equations and response functions}\label{secfluc}

\subsection{Fluctuation equations}

We now follow the well-known path of studying the fluctuations and hydrodynamic response of a holographic system. We perturb the metric and the gauge field by perturbations $\delta g_{\mu\nu}$ and $\delta A_\mu$, respectively:
\begin{eqnarray}
g_{\mu\nu}&\mapsto& g_{\mu\nu}+\delta g_{\mu\nu},\quad \delta g_{\mu\nu}\sim L^2 e^{i k x-i\omega t} h_{\mu\nu}(r),\label{3deltag}\\
A_{\mu}&\mapsto& A_{\mu}+\delta A_{\mu},\quad \delta A_{\mu}\sim \frac{e L}{\kappa} e^{i k x-i\omega t} a_{\mu}(r).\label{3deltaa}
\end{eqnarray}
In this paper we are interested in the transverse (vector) sector, which is expected to contain the diffusion pole.\footnote{The perhaps more interesting longitudinal (scalar) sector, containing the zero sound mode, will be addressed in a separate work. Some results can be found in \cite{Gran:2018jnt}.} We choose the radial gauge: $h_{r\nu}=0$, $a_r=0$. By symmetry, the non-vanishing components are $h_{t y}$, $h_{x y}$ and $a_y$, and the linearized fluctuation equations for them read:\footnote{Although the general form of the fluctuations in Eq.~(\ref{3deltag}-\ref{3deltaa}) is given for $h_{\mu\nu}$ and $a_\mu$, we find it more convenient to work with the fields with one index raised.}
\begin{eqnarray}
    h^{y \prime\prime}_{\phantom{y}t}-\frac{1}{2}\left( \frac{f'}{f}+\frac{g'}{g}+\frac{8}{r} \right) h^{y\prime}_{\phantom{y}t}-\left( k^2 r^2 g+\frac{2r g h \hat{\sigma}}{r\sqrt{f}-1} \right) h^y_{\phantom{y}t}-\omega k r^2 g h^x_{\phantom{x}y}+2 r^2 h' a_y'&=&0,~~~~\label{eoms1}\\
    h^{x\prime\prime}_{\phantom{x}y}+\frac{1}{2}\left( \frac{f'}{f}-\frac{g'}{g}-\frac{4}{r} \right) h^{x\prime}_{\phantom{x}y}+\frac{g}{f}(\omega^2 h^x_{\phantom{x}y}+\omega k h^y_{\phantom{y}t})&=&0,\label{eoms2}\\
    a_y''+\frac{1}{2}\left( \frac{f'}{f}-\frac{g'}{g} \right)a_y'-g \left( k^2 r^2-\frac{\omega^2}{f} \right) a_y+\frac{h'}{r^2 f} h^{y\prime}_{\phantom{y}t}+\frac{g\hat{\sigma}}{r^2\sqrt{f}-r} h^y_{\phantom{y}t}&=&0\label{eoms3}.
\end{eqnarray}
Additionally, there is also a first-order constraint stemming from the gauge choice:
\begin{equation}
    2 r^2 h' a_y+h^{y\prime}_{\phantom{y}t}+\frac{k r^2 f}{\omega} h^{x\prime}_{\phantom{x}y}=0.\label{3constraint}
\end{equation}
However, there is some residual gauge freedom left. Diffeomorphisms and a U(1) transformation generated by a vector $\xi=L^2\xi_{\mu}\mathrm{d}x^{\mu}$ and a scalar $\lambda$ give:
\begin{equation}
    \delta h_{\mu\nu}=-2\nabla_{(\mu}\xi_{\nu)},\quad \delta A_{\mu}=\nabla_{\mu}\lambda.
\end{equation}
Requiring that the radial gauge and the symmetry remain unaffected by these transformations gives $\lambda=0$ and 
$\xi_y\propto 1/r^2$. We get $\delta h_{y t}=i\omega\xi_y$ and $\delta h_{x y}=-i k \xi_y$. Therefore, we can define gauge-invariant linear combinations:
\begin{equation}
    X=k h^y_{\phantom{y}t}+\omega h^x_{\phantom{x}y},\quad Y=a_y,\label{3xy}
\end{equation}
and the system (\ref{eoms1}-\ref{3constraint}) reduces to
\begin{multline}
    X''+\left( \frac{r g h \hat{\sigma}}{2\sqrt{f}}-\frac{2 k^2 r f^2+\omega^2 f'}{f(k^2 r^2 f-\omega^2)}\right) X'-g \left(k^2 r^2-\frac{\omega^2}{f} \right) X\,+\\
    +2 k r^2 h' Y'+2 k\left( r^2\sqrt{f} g \hat{\sigma}-\frac{\omega^2(r^2 f)'h'}{f(k^2 r^2 f-\omega^2)} \right) Y=0,\quad\label{3eomsxy1}
\end{multline}
\begin{equation}
    Y''+\frac{1}{2}\left( \frac{f'}{f}-\frac{g'}{g} \right) Y'+\left( \frac{2\omega^2 {h'}^2}{f(k^2 r^2 f-\omega^2)}-g \left(k^2 r^2-\frac{\omega^2}{f} \right)-\frac{\sqrt{f}g\hat{\sigma}}{h} \right)Y+\frac{k h'}{k^2 r^2 f-\omega^2} X'=0.\label{3eomsxy2}
\end{equation}
These equations can be further simplified by defining a new field:
\begin{equation}
Z=\frac{1}{k^2 r^2 f-\omega^2}\sqrt{\frac{f}{g}}\left( \frac{X'}{r^2}+2k h'Y \right).\label{3zdef}
\end{equation}
The fluctuation equations now become:
\begin{eqnarray}
\left(r^2\sqrt{\frac{f}{g}}Z'\right)'-r^2\left(k^2 r^2 f-\omega^2\right)\sqrt{\frac{g}{f}}Z+2k r^2 h' Y&=&0,\label{3eomszy1}\\
\left(\sqrt{\frac{f}{g}}Y'\right)'-\left(\left(k^2r^2f-\omega^2\right)\sqrt{\frac{g}{f}}+\frac{2{h'}^2}{\sqrt{f g}}+\frac{f\sqrt{g}\hat{\sigma}}{h} \right)Y+k r^2 h' Z&=&0.\label{3eomszy2}
\end{eqnarray}
One final simplification of the equations is possible if we define:
\begin{equation}\label{ZtoU}
U(r)\equiv Z(r)r,
\end{equation}
leading to 
\begin{eqnarray}
\label{3eomsuy1}U''+\frac{1}{2}\left( \frac{f'}{f}-\frac{g'}{g} \right) U'-\left( g \left( k^2r^2-\frac{\omega^2}{f} \right)+\frac{1}{2 r}\left( \frac{f'}{f}-\frac{g'}{g} \right) \right) U+2 k r h' \sqrt{\frac{g}{f}} Y&=&0,\qquad\\
\label{3eomsuy2}Y''+\frac{1}{2}\left( \frac{f'}{f}-\frac{g'}{g} \right)Y'-\left( g \left( k^2r^2-\frac{\omega^2}{f} \right)+\frac{2 {h'}^2}{f}+\frac{\sqrt{f}g\hat{\sigma}}{h} \right) Y+k r h' \sqrt{\frac{g}{f}} U&=&0.\qquad
\end{eqnarray}
At $T=0$ the function $Z$ behaves as $Z\sim 1/r$ in both the IR and the ultraviolet (UV) limit and it turns out very convenient to use $U$ instead of $Z$, since it increases the efficiency of the numerical integration. Yet, at finite temperature the integration of the equations (\ref{3eomszy1}-\ref{3eomszy2}) works faster; therefore, we solve the system (\ref{3eomszy1}-\ref{3eomszy2}) at finite temperature and the system (\ref{3eomsuy1}-\ref{3eomsuy2}) at zero temperature. In Appendix \ref{parity} we show that each of these systems is parity-invariant. Thus we find it sufficient to work only with positive $k$.

\subsubsection{Asymptotic master fields}

One might expect that further transformations of Eqs.~(\ref{3eomszy1}-\ref{3eomszy2}) or (\ref{3eomsuy1}-\ref{3eomsuy2}) could lead to full decoupling of the fluctuation equations in terms of master fields, as in \cite{Edalati:2010hk}, but this is not the case. It is only possible to decouple the equations in the outer region, where we have the RN geometry: in the RN regions the equations are equivalent to those in \cite{Edalati:2010hk}. The master fields in the outer RN region then read:
\begin{equation}\label{3masterfields}
    \Phi_{\pm}=-\frac{k r f}{k^2 r^2 f-\omega^2}X'+c\left( \frac{2k^2 \hat{Q} r^3 f}{k^2 r^2 f-\omega^2}-\frac{3\hat{M}}{2\hat{Q}}\left( 1\pm\sqrt{1+\frac{8k^2\hat{Q}^2}{9\hat{M}^2}} \right) \right)Y,
\end{equation}
and satisfy the equations
\begin{equation}
    (r^2f\Phi_{\pm}')'-c^2\left( k^2-\frac{\omega^2}{r^2f}+2\hat{Q}^2r^2-\frac{3}{2}\hat{M}\left(1\mp\sqrt{1+\frac{8k^2\hat{Q}^2}{9\hat{M}^2}} \right) \right) \Phi_{\pm}=0.
\end{equation}
In order to obtain these results, following \cite{Edalati:2010hk}, we have used the ansatz $\Phi_{\pm}=a(r)Z+b_{\pm}Y$, and found that $a(r)=r$. Then we tried to determine the master fields for the whole bulk in a similar manner: we employed the ansatz $\Phi_{\pm}=a(r)Z+b_{\pm}(r)Y$ (allowing $b_\pm$ to depend on $r$). It turns out that we must set $a(r)=r$ again, since for any other function it must be $b_{\pm}(r)\propto a(r)$, which merely amounts to an overall rescaling of a single function. For the equations to decouple, we find that $b_{\pm}(r)$ must have vanishing first derivatives, i.e.~they must be constant just like in the RN case. However, the expressions we eventually obtain are not constants, and we end up with a contradiction, from which we conclude that the presence of the star makes the decoupling of the equations impossible.\footnote{Although these arguments cannot be considered a firm proof that the decoupling of the equations cannot be done, our conclusion certainly does not contradict what is known about the Kodama-Ishibashi formalism up to date \cite{Jansen:2019wag}.} Consequently, we refer to \eqref{3masterfields} as \emph{asymptotic master fields} (as opposed to true master fields). Although we will not solve them, we will still find them useful when computing the correlation functions from the UV asymptotics.

\subsection{UV asymptotics and Green's functions}\label{secflucuv}

The solution in the UV region is sought for in the form of a power series, as usual with AdS asymptotics. Leaving details for Appendix \ref{UVap}, we note that:
\begin{eqnarray}
    U(r\to0)&=&U^{(0)}+U^{(1)}r+\ldots,\\
    Y(r\to0)&=&Y^{(0)}+Y^{(1)} r+\ldots\enspace .
\end{eqnarray}
The coefficients $U^{(0)}$, $U^{(1)}$, $Y^{(0)}$ and $Y^{(1)}$ are complex constants to be determined from the numerical integration and can easily be expressed in terms of the coefficients that appear in the expansions of $h^x_{\phantom{x}y}$, $h^y_{\phantom{y}t}$ and $a_y$.

Since the far UV region certainly belongs to the outer RN part of the background solution, it allows us to use the asymptotic master fields from Eq.~(\ref{3masterfields}), which also have the asymptotic expansion:
\begin{equation}
    \Phi_{\pm}(r\to0)=\Phi_{\pm}^{(0)}+\Phi_{\pm}^{(1)}r+\ldots
\end{equation}
Following \cite{Edalati:2010hk}, we use these fields to define:
\begin{equation}
    G_{\pm}(\omega, k)=\left(\frac{1}{\sqrt{1+\frac{8k^2\hat{Q}^2}{9\hat{M}^2}} }\pm 1\right)\frac{\Phi^{(1)}_{-}}{\Phi^{(0)}_{-}}-\left(\frac{1}{\sqrt{1+\frac{8k^2\hat{Q}^2}{9\hat{M}^2}}}\mp 1\right)\frac{\Phi^{(1)}_{+}}{\Phi^{(0)}_{+}}.\label{3gpm}
\end{equation}
We will refer to $G_{\pm}$ as to the \emph{auxiliary Green's functions}. We can express $G_{\pm}$ either as functions of the gauge-dependent coefficients of $h^x_{\phantom{x}y}$, $h^y_{\phantom{y}t}$ and $a_y$, or as functions of the numerically obtained gauge-invariant quantities $U^{(0)}$, $U^{(1)}$, $Y^{(0)}$ and $Y^{(1)}$. In the latter case they read:
\begin{eqnarray}
    G_+=\frac{3\hat{M} Y^{(0)}U^{(1)}+\hat{Q} k\left(U^{(0)}U^{(1)}-2Y^{(0)}Y^{(1)} \right)}{3\hat{M} Y^{(0)}U^{(0)}+\hat{Q}k\left({U^{(0)}}^2-2{Y^{(0)}}^2 \right)},\label{G+}\\
    G_-=-\frac{3\hat{M} Y^{(1)}U^{(0)}+\hat{Q} k\left(U^{(0)}U^{(1)}-2Y^{(0)}Y^{(1)} \right)}{3\hat{M} Y^{(0)}U^{(0)}+\hat{Q}k\left({U^{(0)}}^2-2{Y^{(0)}}^2 \right)}.\label{G-}
\end{eqnarray}
In this way, we are able to express any gauge-dependent response in terms of (numerically calculated) $G_{+}$ and $G_{-}$ and its sources. Since we are dealing with operator mixing, this proves to be a necessary step when we calculate the boundary action. Note that $G_{\pm}$ are parity-invariant.

We determine the renormalized, on-shell boundary action and hence the field-theory correlation functions of $T_{yt}, T_{xy}$ and $J_y$ using the prescription from \cite{Kaminski:2009dh}. Neglecting contact terms, we find:
\begin{eqnarray}
    G_{ytyt}(\omega, k)&=&-\frac{k^2}{2c}G_+(\omega, k),\label{3gytyt}\\
    G_{ytxy}(\omega, k)&=&\frac{\omega k}{2c}G_+(\omega, k),\label{3gytxy}\\
    G_{xyxy}(\omega, k)&=&-\frac{\omega^2}{2c}G_+(\omega, k),\label{3gxyxy}\\
    G_{yty}(\omega, k)&=&\frac{\hat{Q}}{3\hat{M}}k^2\left(G_+\left(\omega, k\right)+G_-\left(\omega, k\right)\right),\label{3gyty}\\
    G_{xyy}(\omega, k)&=&-\frac{\hat{Q}}{3\hat{M}}\omega k\left(G_+\left(\omega, k\right)+G_-\left(\omega, k\right)\right),\label{3gxyy}\\
    G_{yy}(\omega, k)&=&c G_-(\omega, k).\label{3gyy}
\end{eqnarray}
The dimensionless long-wave transverse conductivity and the dimensionless viscosity are then given by the following relations \cite{Hartnoll:2010gu, Hartnoll:2010ik}:
\begin{equation}
    \hat{\sigma}_y(\omega)\equiv e^2\sigma_y(\omega)=\frac{i c}{\omega}G_{yy}(\omega, 0),\qquad \hat{\eta}\equiv\frac{\kappa^2}{L^2}\eta=-\lim_{\omega\to 0}\frac{1}{\omega}\text{Im}\,G_{xyxy}(\omega, 0).
\end{equation}
Since the dimensionless entropy density at finite temperature reads $\hat{s}=2\pi$ \cite{Hartnoll:2010ik}, we expect to obtain $\hat{\eta}=1/2$ at finite temperature.

We are now ready to perform the calculations: we have specified the equations of motion, we have found the UV boundary conditions which also specify the prescription for the correlators, while the IR boundary conditions are found from the deep interior or near-horizon expansion at $T=0$ and $T>0$, respectively -- we give them in Appendix \ref{secflucir}. Numerical calculations of the fluctuations are in principle straightforward: we solve an initial value problem in \verb|Mathematica| using the command \verb|NDSolve|. We set the boundary condition at the horizon at finite $T$ and in deep IR at $T=0$ (introducing a large but finite IR cutoff). In both cases we normalize the IR solutions to unity.\footnote{The overall normalization of solutions is usually unimportant; however, we will see in Section \ref{analytical green's functions} that the normalization of $Y$ turns out crucial for understanding of its behavior near the boundary.}

There are, however, numerous practical difficulties with stability and convergence. In the first place, it is important to use the maximally simplified equations: (\ref{3eomszy1}-\ref{3eomszy2}) at finite $T$ and (\ref{3eomsuy1}-\ref{3eomsuy2}) at $T=0$, because of the factor $k^2r^2f-\omega^2$, which appears in denominators of the system (\ref{3eomsxy1}-\ref{3eomsxy2}), leading to singularities (for every real $\omega$ and $k$ there is a point $r$ such that this factor equals zero). Furthermore, one needs to carefully adjust the precision of the numerics. For example, at finite $T$ we had to set \verb|WorkingPrecision| $\to30$. On the other hand, at $T=0$ it is enough to set \verb|WorkingPrecision| $\to12$, but the frequencies must not be too high: for $\omega$ too large the solutions undergo rapid oscillations throughout the bulk, becoming indistinguishable from noise.

\section{Results: correlation functions and hydrodynamics}\label{secresults}

In order to compute the correlation functions for a given species of fermions in the bulk, we solve the equations of motion for temperatures ranging from $T=0$ to $T=T_c$ for a fixed $\hat{\beta}$. Keeping $\hat{\beta}$ fixed ensures that we deal with the same kind of fermions at all temperatures, while a systematic temperature increase accounts to tracing the system from the pure electron star to a thin shell, eventually supposed to disappear at $T_c$ in favor of the RN black hole. Practically, this means that we choose the mass $\hat{m}$ and the Lifshitz exponent $z$ at $T=0$, calculate $\hat{\beta}$, and keep using this $\hat{\beta}$ at all temperatures. Different choices of $z$ then yield different families of solutions (sourced by different species of fermions). We are interested in small, intermediate and large values of $z$, where the last case is of particular interest since the limit $z\to\infty$ corresponds to the RN metal.

We have found it most convenient to fix $\hat{m}=0.1$ and work with $z=2.5$, $z=4$ and $z=20$ at $T/T_c\in \{0, 0.05, 0.23, 0.35, 0.65, 0.95 \}$. Numerical integration of the background equations of motion (\ref{2eom1}-\ref{2eom3}) proves to be challenging for some parameter values: we were unable to generate the zero-temperature background for $z=20$, while the lowest finite temperatures we were able to reach are $T/T_c=0.35$ for $z=2.5$, $T/T_c=0.23$ for $z=4$ and $T/T_c=0.05$ for $z=20$. Although we may not have enough information on the limit $T\to0$, the insight from the remaining temperatures is nevertheless extremely rich. As a sanity check, we have calculated the viscosity coefficient for all finite temperatures, obtaining $\hat{\eta}=0.5$ (i.e.~$\hat{\eta}/\hat{s}=1/4\pi$) to high accuracy, as expected.

\begin{figure}[t]
    \centering
    \includegraphics[width=\linewidth]{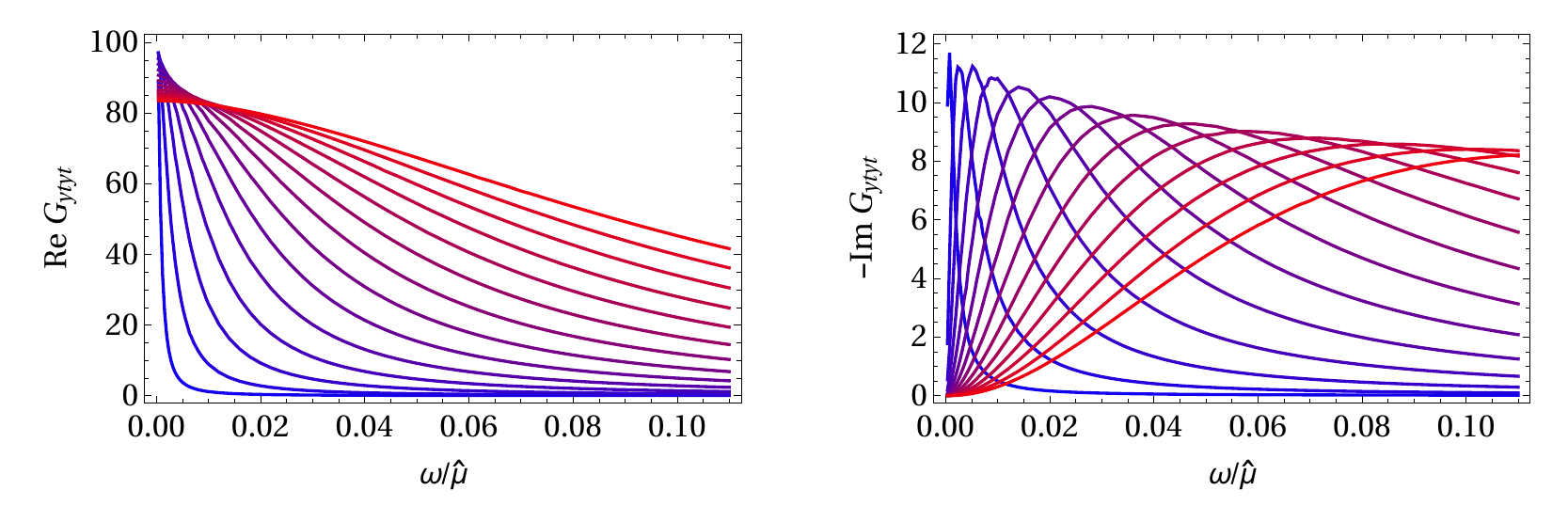}\\
    \includegraphics[width=\linewidth]{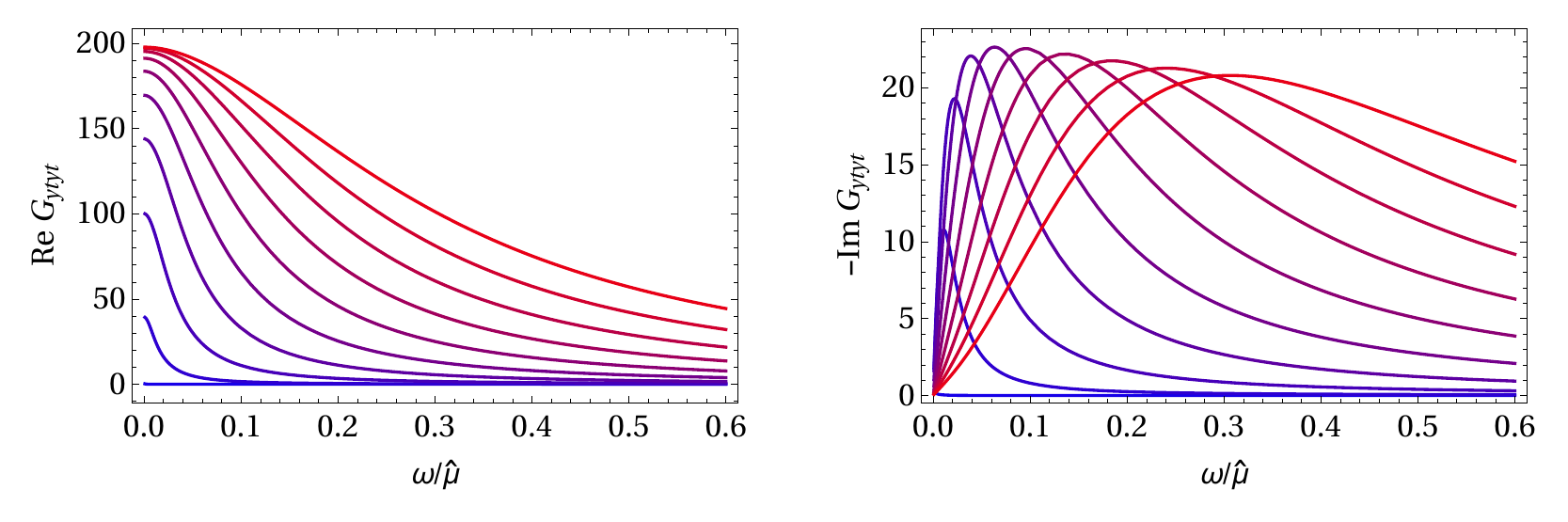}\vspace{-0.5cm}
    \caption{The real part (left) and the negative imaginary part (right) of $G_{ytyt}$ for $z=4$, for $T/T_c=0$ and a range of momenta $k/\hat{\mu}\in[0.01,0.14]$ (from blue to red; top) and for $T/T_c=0.23$ and a range of momenta $k/\hat{\mu}\in[0.001,0.091]$ (from blue to red; bottom). For these temperatures we see the typical diffusion peak in $\mathrm{Im}\,G_{ytyt}$.}
    \label{linplot4_0_023}
\end{figure}
\begin{figure}[t]
    \centering
    \includegraphics[width=\linewidth]{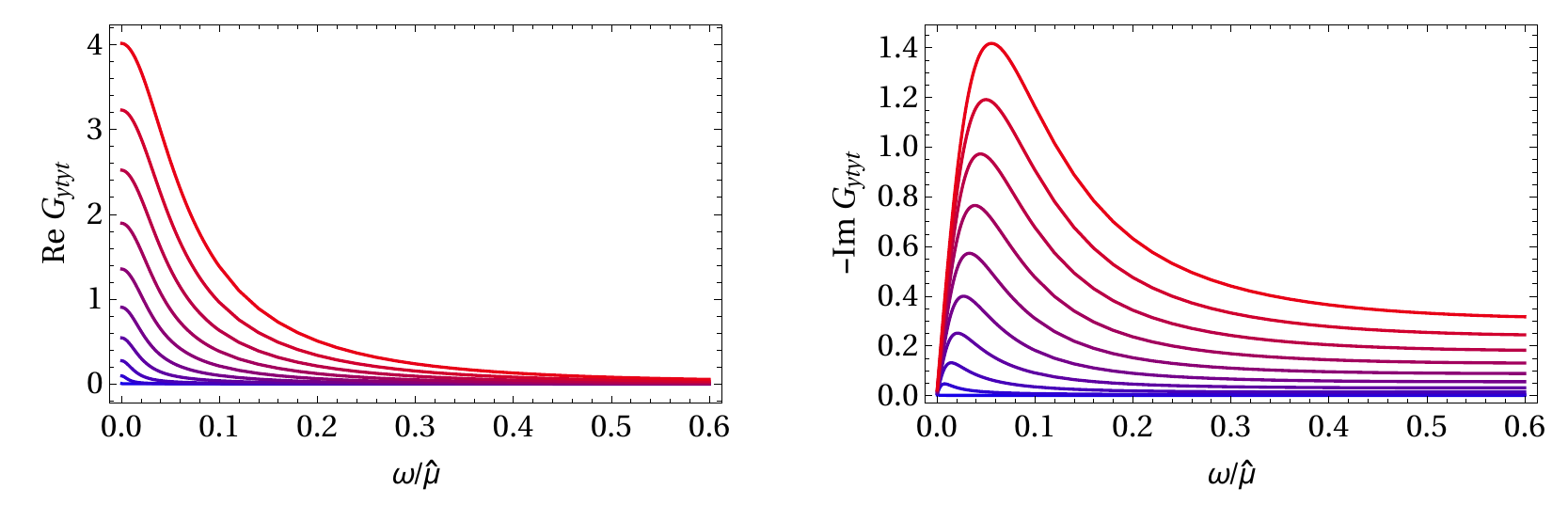}\\
    \includegraphics[width=\linewidth]{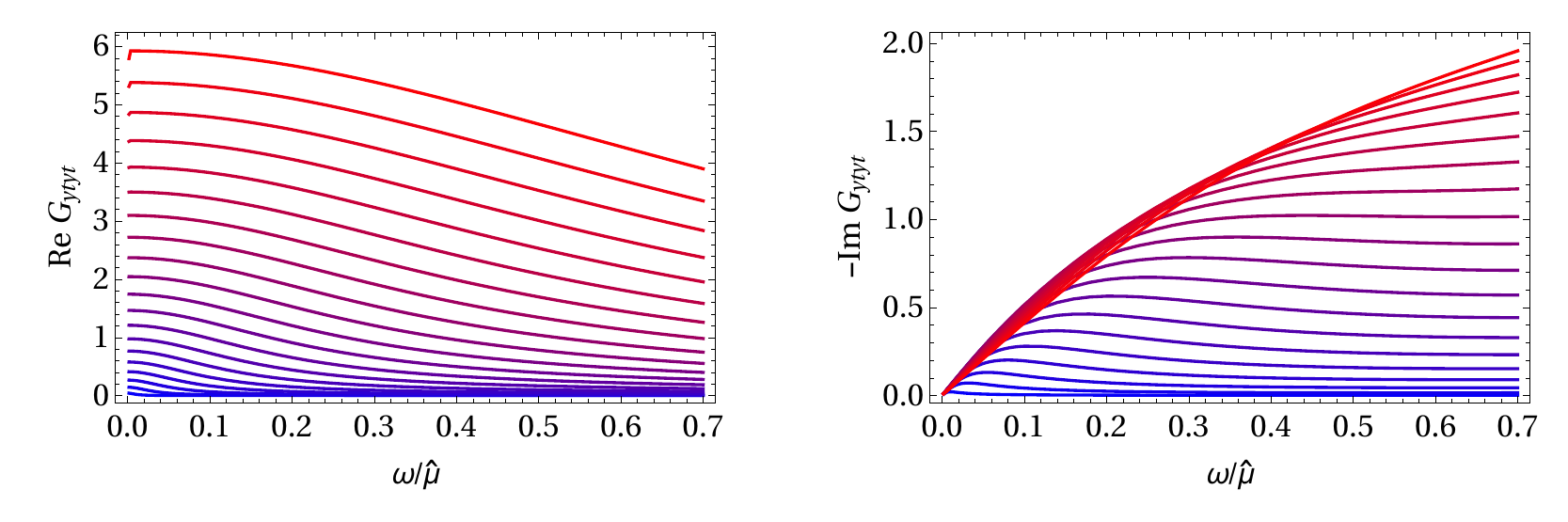}\vspace{-0.5cm}
    \caption{The real part (left) and the negative imaginary part (right) of $G_{ytyt}$ for $z=4$, for $T/T_c=0.35$ and a range of momenta $k/\hat{\mu}\in[0.03,0.6]$ (from blue to red; top) and for $T/T_c=0.65$ and a range of momenta $k/\hat{\mu}\in[0.02,0.78]$ (from blue to red; bottom). In the top panels the typical form of a diffusive correlator is visibly deformed. In the bottom panels the diffusion pole vanishes completely.}
    \label{linplot4_035_065}
\end{figure}
\begin{figure}[t]
    \centering
    \includegraphics[width=\linewidth]{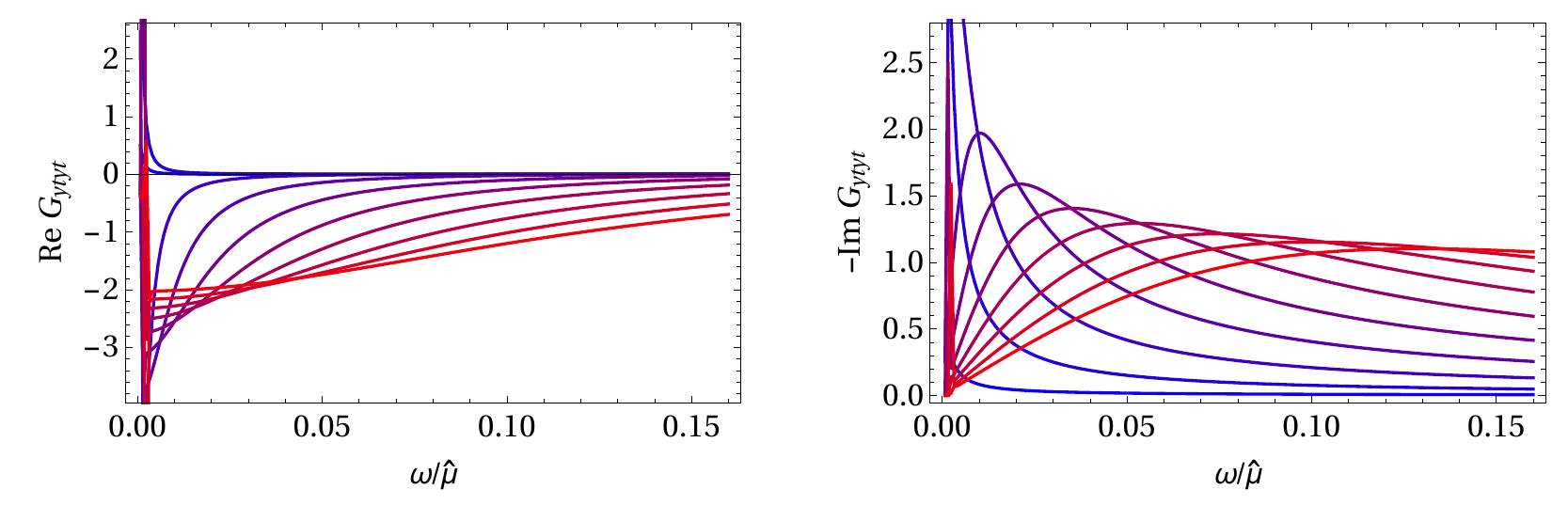}\\
    \includegraphics[width=\linewidth]{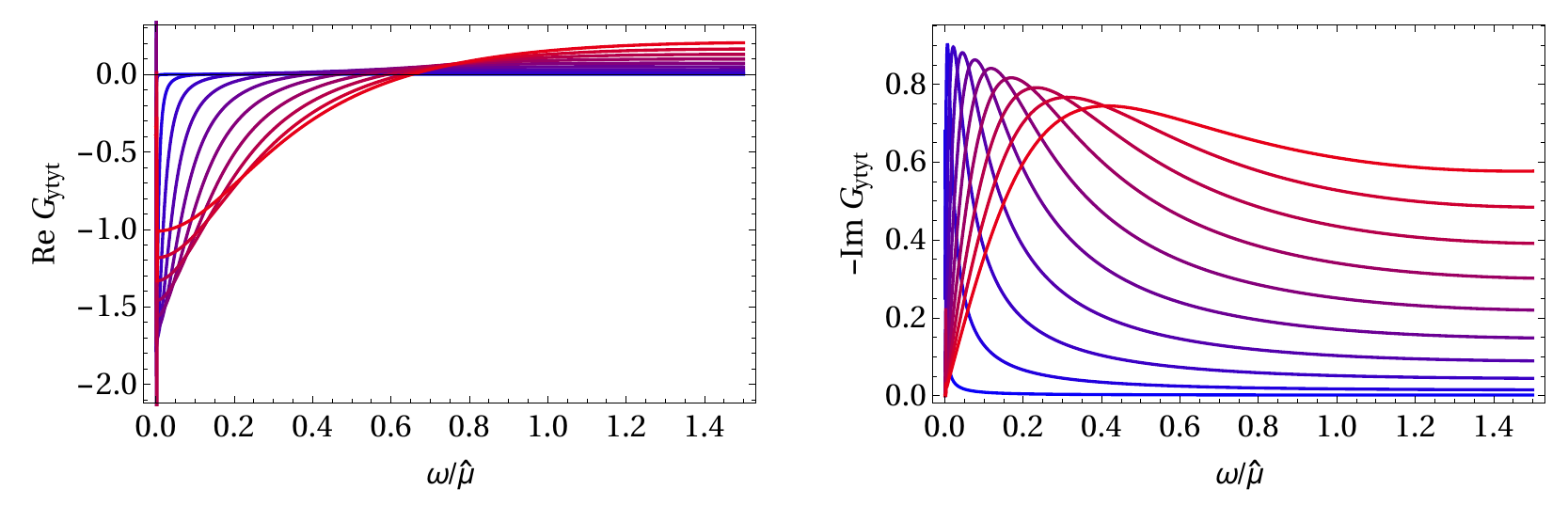}\vspace{-0.5cm}
    \caption{The real part (left) and the negative imaginary part (right) of $G_{ytyt}$ for $z=4$ and $T/T_c=0.95$, for a range of momenta $k/\hat{\mu}\in[0.05,0.95]$ (from blue to red; top) and for the RN case at $T/T_c=0.95$, for a range of momenta $k/\hat{\mu}\in[0.05,1.4]$ (from blue to red; bottom). At high temperature the hydrodynamic diffusion is restored.}
    \label{linplot4_095_RN}
\end{figure}

\begin{figure}[t]
    \begin{minipage}{0.5\textwidth}
        \centering
        \includegraphics[height=0.28\textheight]{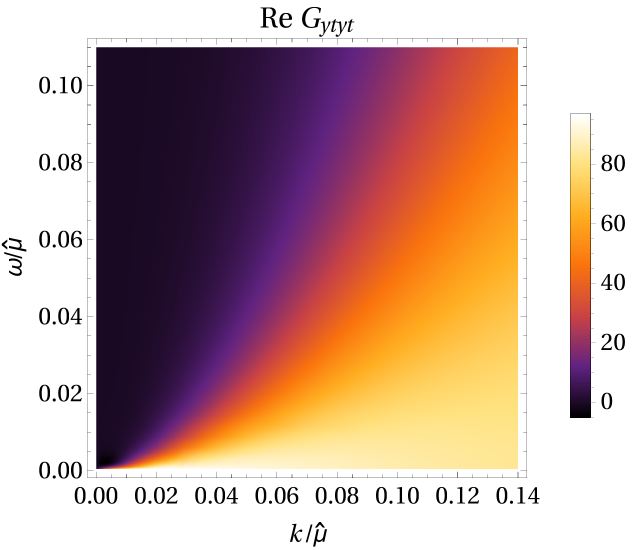}
    \end{minipage}
    \begin{minipage}{0.5\textwidth}
        \centering
        \includegraphics[height=0.28\textheight]{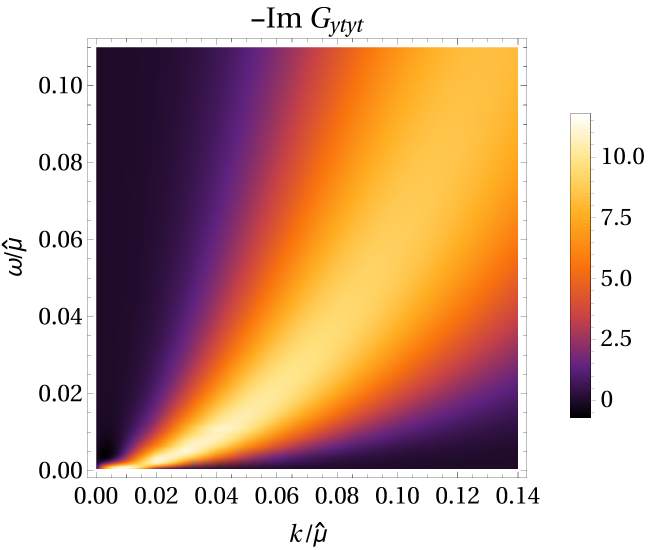}
    \end{minipage}\vspace{5pt}\\
    \begin{minipage}{0.5\textwidth}
        \centering
        \includegraphics[height=0.28\textheight]{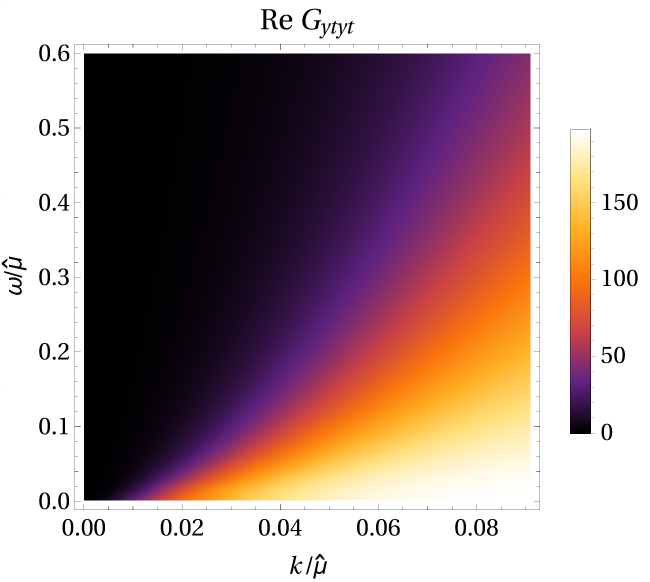}
    \end{minipage}
    \begin{minipage}{0.5\textwidth}
        \centering
        \includegraphics[height=0.28\textheight]{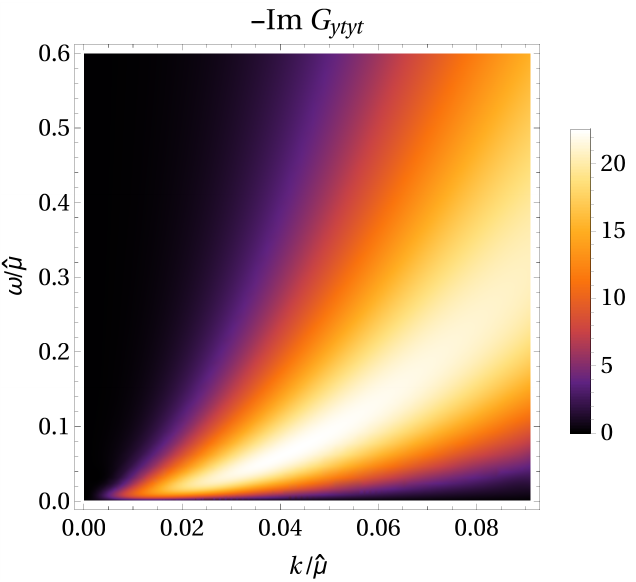}
    \end{minipage}
    \caption{Density plots of the real part (left) and the negative imaginary part (right) of $G_{ytyt}$ for $z=4$, for $T/T_c=0$ (top) and $T/T_c=0.23$ (bottom). The diffusive behavior $\omega\propto -ik^2$ is obvious (compare to Figure~\ref{linplot4_0_023}).}
    \label{dplot4_0_023}
\end{figure}
\begin{figure}[t]
    \begin{minipage}{0.5\textwidth}
        \centering
        \includegraphics[height=0.28\textheight]{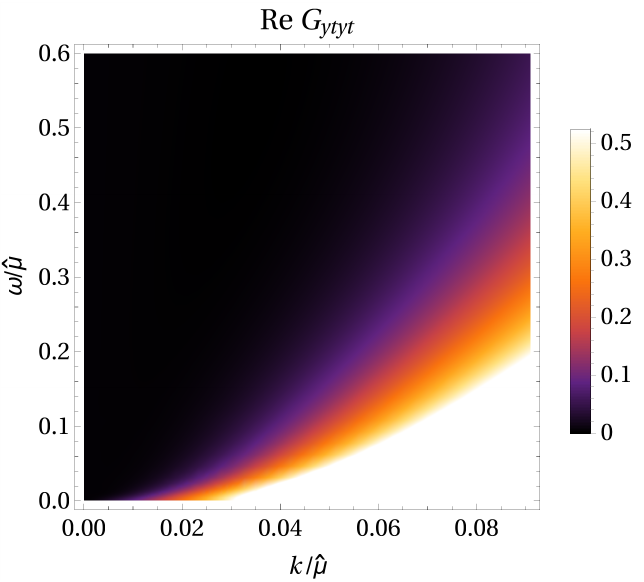}
    \end{minipage}
    \begin{minipage}{0.5\textwidth}
        \centering
        \includegraphics[height=0.28\textheight]{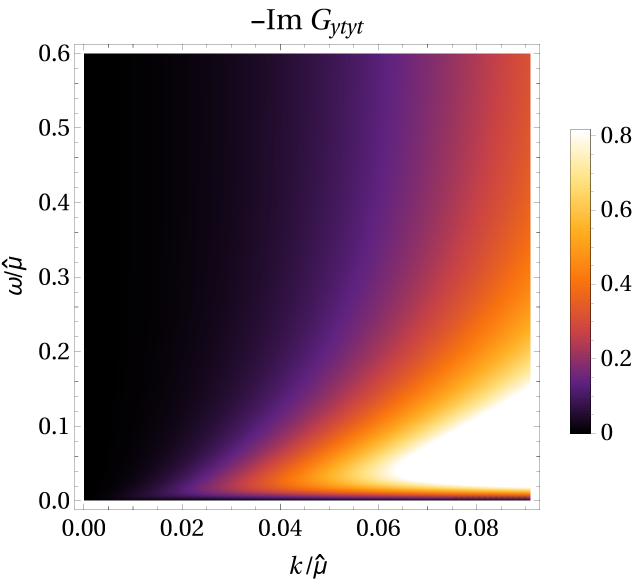}
    \end{minipage}\vspace{5pt}\\
    \begin{minipage}{0.5\textwidth}
        \centering
        \includegraphics[height=0.28\textheight]{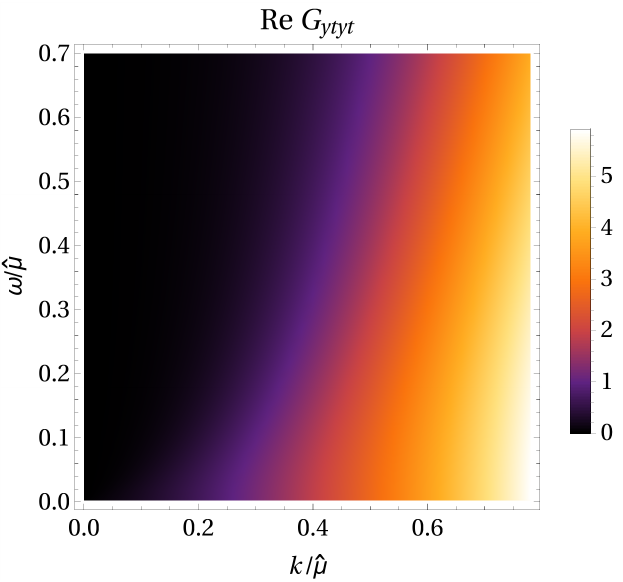}
    \end{minipage}
    \begin{minipage}{0.5\textwidth}
        \centering
        \includegraphics[height=0.28\textheight]{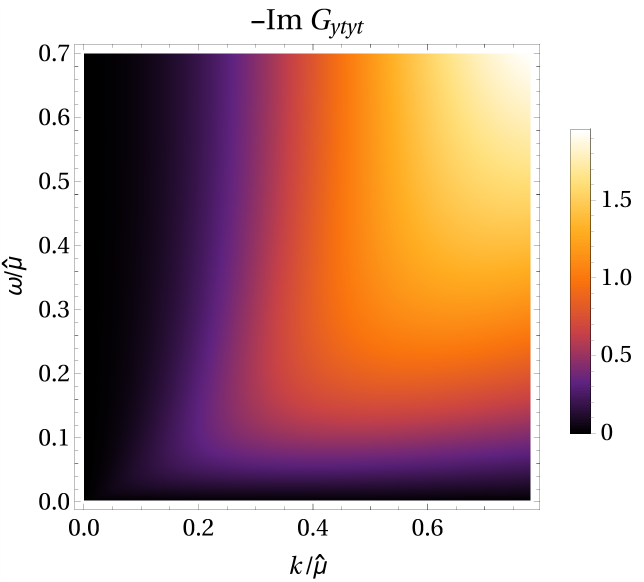}
    \end{minipage}
    \caption{Density plots of the real part (left) and the negative imaginary part (right) of $G_{ytyt}$ for $z=4$, for $T/T_c=0.35$ (top) and $T/T_c=0.65$ (bottom). Just like in the linear plots in Figure~\ref{linplot4_035_065}, the diffusive behavior disappears.}
    \label{dplot4_035_065}
\end{figure}
\begin{figure}[t]
    \begin{minipage}{0.5\textwidth}
        \centering
        \includegraphics[height=0.28\textheight]{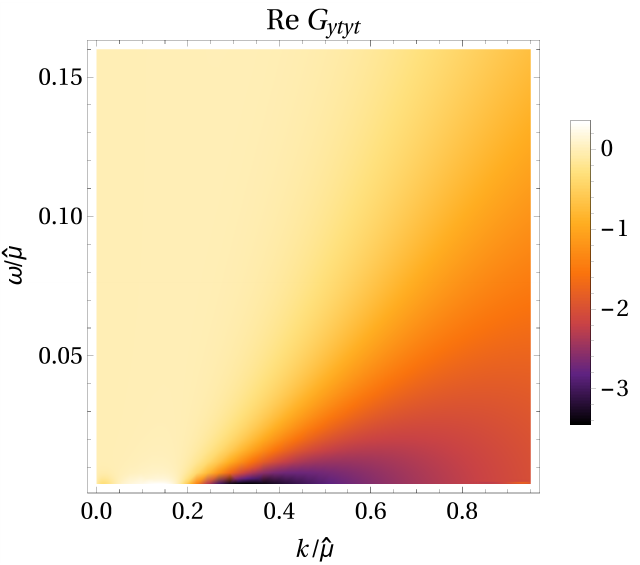}
    \end{minipage}
    \begin{minipage}{0.5\textwidth}
        \centering
        \includegraphics[height=0.28\textheight]{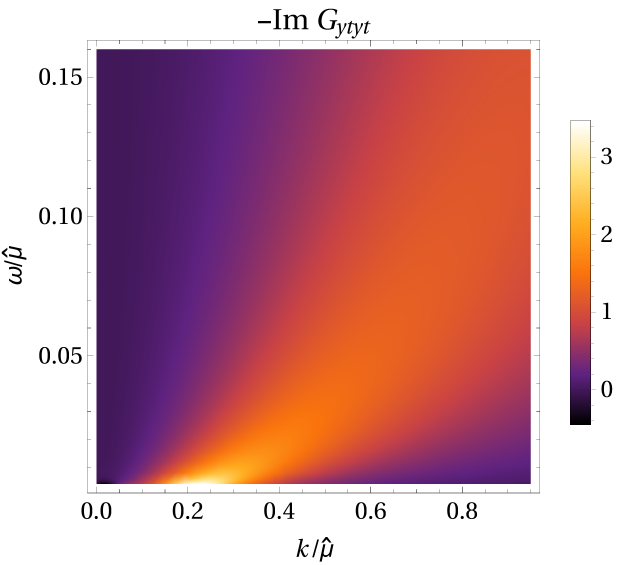}
    \end{minipage}\vspace{5pt}\\
    \begin{minipage}{0.5\textwidth}
        \centering
        \includegraphics[height=0.28\textheight]{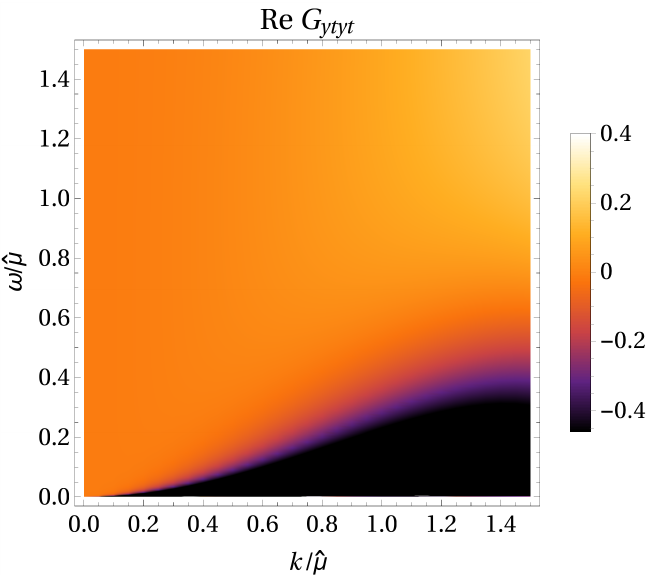}
    \end{minipage}
    \begin{minipage}{0.5\textwidth}
        \centering
        \includegraphics[height=0.28\textheight]{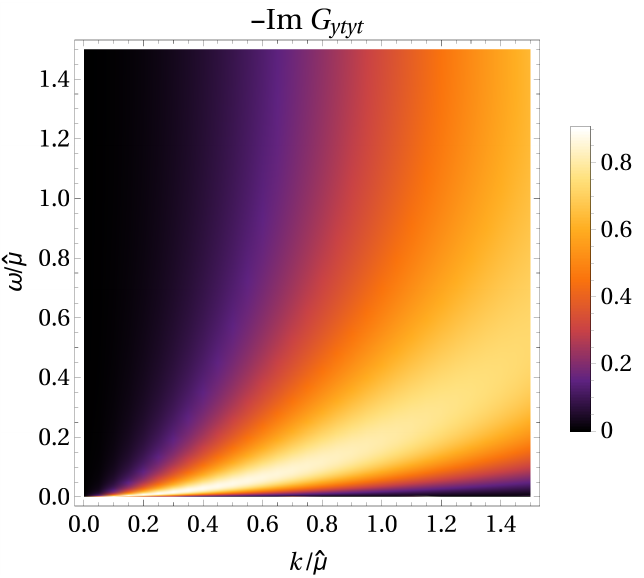}
    \end{minipage}
    \caption{Density plots of the real part (left) and the negative imaginary part (right) of $G_{ytyt}$ for $z=4$ and $T/T_c=0.95$ (top) and for the RN case at $T/T_c=0.95$ (bottom); hydrodynamics re-emerges, cf. Figure~\ref{linplot4_095_RN}.}
    \label{dplot4_095_RN}
\end{figure}

Apart from the single exception of $z=20$ at $T/T_c=0.05$,\footnote{This case exhibits very peculiar features and generally does not follow any of the trends common to the rest of the solutions. It might be that such behavior emerges when one approaches the zero temperature in general, but for now we are not able to give any definite answer to this puzzle.} similar results are obtained for all three $z$ values so we choose to present the results for $z=4$. Since all the correlators depending on $G_+$ exhibit similar properties, we present only $G_{ytyt}$. Figures~\ref{linplot4_0_023}-\ref{linplot4_095_RN} show its frequency dependence for various momenta and five different temperatures: $T/T_c=0$, $T/T_c=0.23$ (Figure \ref{linplot4_0_023}), $T/T_c=0.35$, $T/T_c=0.65$ (Figure \ref{linplot4_035_065}) and $T/T_c=0.95$ (Figure \ref{linplot4_095_RN}).\footnote{The imaginary parts are given with a negative sign since $-\mathrm{Im}\,G_{ytyt}$ is always positive.} The bottom row in Figure \ref{linplot4_095_RN} is again for $T/T_c=0.95$ but for the RN black hole for comparison. The corresponding $k$-$\omega$ density plots are shown in Figures~\ref{dplot4_0_023}-\ref{dplot4_095_RN} (for these figures we sometimes use denser momentum grids in order to show the full 2D energy-momentum dependence). We express $\omega$ and $k$ in the units of the field theory chemical potential.\footnote{There are two scales in our theory: the field theory chemical potential $\hat{\mu}$ and the temperature $T$. We have found the former more convenient to work with.} 

The data shown suggest the following conclusions:
\begin{enumerate}
\item At zero and low temperatures (Figures \ref{linplot4_0_023} and \ref{dplot4_0_023}) we recognize the characteristic form of a diffusion correlator, i.e.~hydrodynamic response.
\item At high temperatures (Figures \ref{linplot4_095_RN} and \ref{dplot4_095_RN}) we likewise see nice hydrodynamic diffusion, just like for the RN case which is known to be hydrodynamic \cite{Edalati:2010hk,Davison:2013bxa}.
\item At intermediate temperatures (Figures \ref{linplot4_035_065} and \ref{dplot4_035_065}) the correlation functions look very different and clearly cannot be described by (normal) hydrodynamics.
\end{enumerate}
It is not that difficult to understand the above points on a qualitative level. At low temperatures, the horizon is small (at $T=0$ it is nonexistent) and the response of the system is dominated by the electron star. The fact that it is a fluid system, that it has metallic conductivity as found in \cite{Hartnoll:2010gu} and that it is approximately described by Lifshitz geometry in IR, which is known to exhibit hydrodynamic behavior at finite (but arbitrarily small) temperatures at least for $z>2$ \cite{Sybesma:2015oha,Gursoy:2016tgf} all suggest that it could have a hydrodynamic response. At high temperatures the horizon is large and the electron cloud is small, hence the system should not deviate much from the RN hydrodynamics. The RN hydrodynamics is very well studied \cite{Davison:2013bxa}, and this is why we make a comparison between the electron cloud and the pure RN black hole at $T/T_c=0.95$. The comparison is handily eased by the fact that in the pure RN case $\hat{\mu}=\hat{\mu}_0\approx0.65$, while in the presence of the cloud $\hat{\mu}\approx0.67$.\footnote{Another way to compare the two geometries is to recast Eq.~\eqref{3eomszy2} into a Schr\"odinger form by setting $Z=0$ and analyze the resulting effective potential. The effective potentials in the two cases agree in the asymptotic regions (in the IR they are the same), but may differ considerably in the region occupied by the electron cloud. In general, the effective potential in the presence of the cloud is deeper and admits more bound states, which agrees with the core definition of the electron star.}

At intermediate temperatures, both subsystems (the horizon and the cloud) are non-negligible and their interplay drastically changes the picture. It looks as if the diffusion pole moves towards higher frequencies and momenta as the temperature increases (until it increases so much that we re-enter the hydrodynamic regime). For now, this is just a handwaving claim based on the visual inspection of the figures; in Section \ref{analytical green's functions} we will corroborate this analytically. In any case, there is no sign of a hydrodynamic regime at intermediate $T/T_c$. In fact, things are even more bizarre at intermediate temperatures. In the next section we will see that this state is unstable -- a pole develops in the upper half of the complex frequency plane, signaling that the electron cloud solution might be a false vacuum in this regime.

\begin{figure}[t]
    \includegraphics[width=0.975\linewidth]{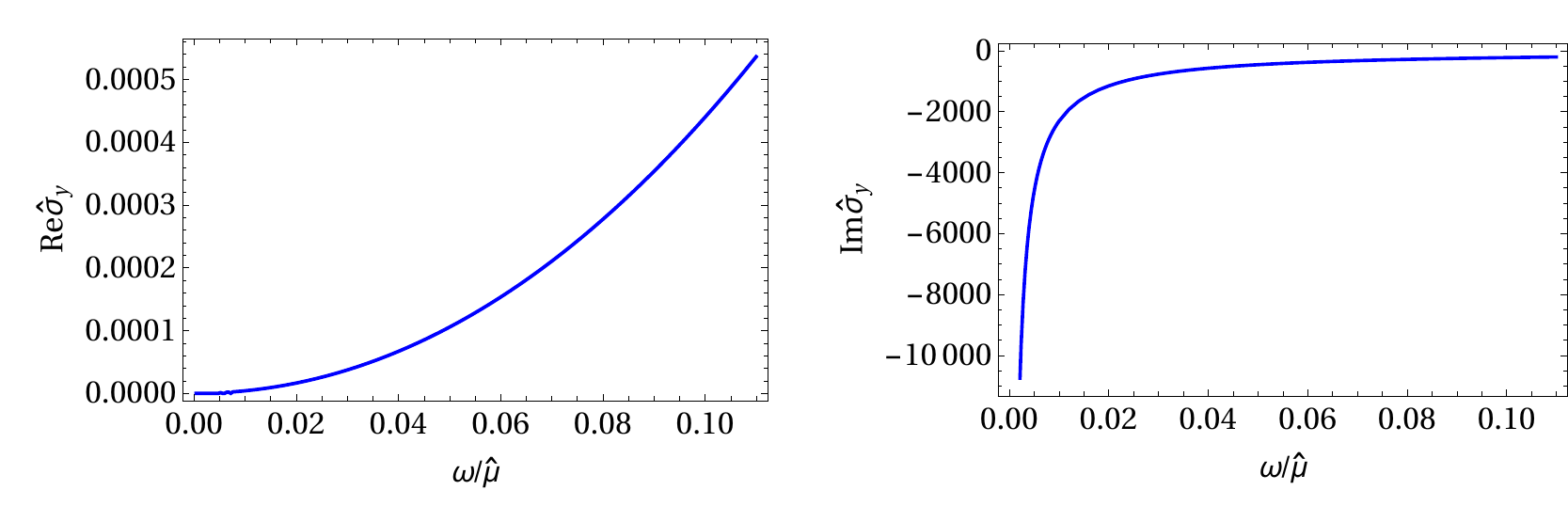}\\
    \includegraphics[width=\linewidth]{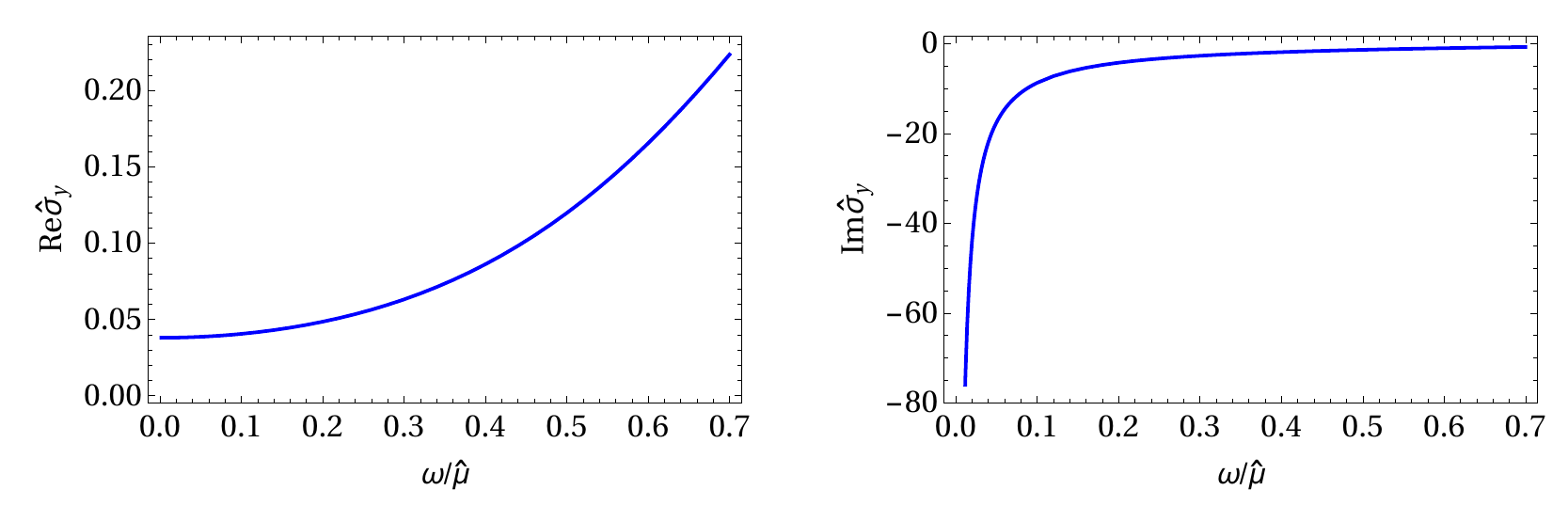}\vspace{-0.2cm}
    \caption{The real part (left) and the imaginary part (right) of the transverse conductivity $\hat{\sigma}_y$ for $z=4$, for $T/T_c=0$ (top) and $T/T_c=0.65$ (bottom). The conductivity is rather featureless, as it depends solely on $G_-$.}
    \label{condplot}
\end{figure}

Transverse conductivity $\hat{\sigma}_y$ is given in Figure~\ref{condplot}. It has a similar profile at all temperatures and for all Lifshitz exponents. The only difference is the value of $\mathrm{Re}\,\hat{\sigma}_y$ in the limit $\omega\to0$, which grows with temperature for every $z$.\footnote{It is understood that extending $\mathrm{Re}\,\hat{\sigma}_y$ to $\omega=0$ gives a $\delta$-function.} Otherwise, the conductivity is mainly featureless. Specifically, there is no sign of a diffusion or any other pole at low frequencies, so we infer that $G_-$ (see Eq.~\eqref{3gyy}) is a holomorphic function in the entire complex frequency plane. In Section \ref{analytical green's functions} we will confirm analytically that this is true at least in the low-energy limit.

\section{Semianalytical low-energy expansion}\label{secanal}

It is possible to solve the equations of motion in the limit of small $\omega$ and $k$ at finite temperature following the lines of \cite{Policastro:2002se,Policastro:2002tn,Kovtun:2005ev,Starinets:2008fb}. As we are going to see, a fully analytical approach is impossible in the presence of the cloud, meaning that we still have to find the boundary values of $Y$ numerically. Nevertheless, we will be able to determine closed-form expressions for $Z$ and $Y$ and get insight into the structure of Green's functions. 

We start by switching from the radial coordinate $r$ to the tortoise coordinate $r_*$, defined as:
\begin{equation}
r_*(r)=\int^r_0 \sqrt{\frac{g(r')}{f(r')}}\mathrm{d}r'.
\end{equation}
In the UV limit we have $r_*\approx r/c$, while in the IR limit:
\begin{eqnarray}
    r_*(r\to 1)&\approx& -\frac{\log (1-r)}{4\pi c T},\quad T>0,\\
    r_*(r\to\infty)&\approx& \frac{\sqrt{g_{\infty}}}{z}r^z,\quad T=0.
\end{eqnarray}
Next, we factor out the singular parts of the infalling solutions at the horizon:
\begin{equation}
    Z(r_*)=\zeta(r_*)e^{i\omega r_*},\quad Y(r_*)=\upsilon(r_*)e^{i\omega r_*},
\end{equation}
so that the equations of motion become
\begin{eqnarray}
    (r^2\zeta')'+2i\omega r (r\zeta)'-k^2r^4f\zeta+2k r^2 h'\upsilon&=&0,\label{Aeoms1}\\
    \upsilon''+2i\omega\upsilon'-(V+k^2r^2f)\upsilon+k r^2 h'\zeta&=&0,\label{Aeoms2}
\end{eqnarray}
where for clarity we define
\begin{equation}
    V=\frac{2{h'}^2}{f}+\frac{f^{3/2}\hat{\sigma}}{h},
\end{equation}
and where all the derivatives are with respect to $r_*$, while $r$ is understood to be a function of $r_*$.

Assume now that the equations and their solutions can be expanded in a small parameter $\epsilon$, so that
\begin{eqnarray}
    \zeta(r_*)&=&\zeta_0(r_*)+\zeta_1(r_*)+\zeta_2(r_*)+\ldots,\label{eq58}\\
    \upsilon(r_*)&=&\upsilon_0(r_*)+\upsilon_1(r_*)+\upsilon_2(r_*)+\ldots,\label{eq59}
\end{eqnarray}
with $\zeta_n\sim\upsilon_n\sim\epsilon^n$ for an integer $n\ge 0$. We assume that $k\sim\epsilon$ and $\omega\sim\epsilon^{\gamma}$ for some real parameter $\gamma$, $\gamma\ge1$. 
In this way we allow that $\omega\sim |k|^{\gamma}$ need not satisfy the normal diffusive dispersion relation. However, this immediately poses a difficulty: because of the scaling $\omega\sim\epsilon^\gamma$ we cannot know in advance at which order in the $\epsilon$-expansion the $\omega$-dependent terms should appear. These terms may be of order $\epsilon$, $\epsilon^2$, or somewhere ``in-between'' -- $\gamma$ is not known in advance and need not be an integer. But this will turn out irrelevant as we will eventually see, since the structure of the equations allows the leading $\omega$-term to be captured. Thus, we expand the equations assuming first that $\omega\sim\epsilon$. Then to second order we get:
\begin{eqnarray}
    (r^2\zeta_0')'&=&0,\label{zetaeq0}\\
    (r^2\zeta_1')'&=&-2k r^2h'\upsilon_0-i\omega(r^2\zeta_0'+(r^2\zeta_0)'),\label{zetaeq1}\\
    (r^2\zeta_2')'&=&k^2r^4f\zeta_0-2k r^2 h'\upsilon_1-i\omega(r^2\zeta_1'+(r^2\zeta_1)')\label{zetaeq2}
\end{eqnarray}
and
\begin{eqnarray}
    \upsilon_0''-V\upsilon_0&=&0,\label{yeq0}\\
    \upsilon_1''-V\upsilon_1&=&-k r^2 h'\zeta_0-2i\omega\upsilon_0',\label{yeq1}\\
    \upsilon_2''-V\upsilon_2&=&k^2r^2f\upsilon_0-2i\omega\upsilon_1'-k r^2h'\zeta_1.\label{yeq2}
\end{eqnarray}
The equations (\ref{zetaeq0}-\ref{zetaeq2}) have a nice property that they can be solved by quadratures -- this is the advantage of switching to the tortoise coordinate. We obtain $\zeta$ by integrating these equations twice and by choosing the integration constants in such a way that the solutions are regular at the horizon (i.e.~so that the logarithmic divergences at the horizon cancel), and assuming that $\upsilon(r_*\to\infty)\approx\mathrm{const}$. The result is
\begin{equation}\label{zetasol}
   \zeta(r_*)= \left[ 1+\int^{\infty}_{r_*}\left( -i\omega+\int^{\infty}_{r_*'} (k^2r^4f-2|k| r^2(\upsilon_0+\upsilon_1)h')\mathrm{d}r_*'' \right)\frac{\mathrm{d}r_*'}{r^2}\right]\frac{k}{|k|}+\mathcal{O}(\omega k).
\end{equation}
We normalize the solution to unity at the horizon (up to the sign of $k$) in order to match the boundary conditions for the numerical integration, conserving the parity (see Appendix \ref{parity}).\footnote{We assume that $k\neq0$.}

Equations (\ref{yeq0}-\ref{yeq2}) can also be easily solved provided that we know two independent solutions to Eq.~\eqref{yeq0} (let us call them $\upsilon_0^{(1)}$ and $\upsilon_0^{(2)}$), since the homogeneous parts of all three equations have the same form. However, as we mentioned above, these solutions can only be found numerically. In order to distinguish the two numerical solutions, we set independent initial conditions at the horizon. We find:
\begin{eqnarray}
    \upsilon_0^{(1)}(r_*\to\infty)&=&I_0\left( \hat{\mu}_0\sqrt{\frac{2}{\pi c T}}e^{-2\pi c T r_*} \right),\\
    \upsilon_0^{(2)}(r_*\to\infty)&=&K_0\left( \hat{\mu}_0\sqrt{\frac{2}{\pi c T}}e^{-2\pi c T r_*} \right),
\end{eqnarray}
where $I_0$ and $K_0$ are the modified Bessel functions (the $I_0$ branch equals unity at the horizon and agrees with the IR expansion used as the initial condition for the numerics, cf. Eq.~\eqref{irexp}). Again, we determine the constants of integration demanding that the solution at the horizon is regular and correctly normalized (equal to unity in our conventions). We obtain:
\begin{multline}
    \upsilon(r_*)=\upsilon_0^{(1)}-\upsilon_0^{(1)}\int^{\infty}_{r_*}  \frac{\upsilon_0^{(2)}}{W_{1,2}}   \left( 2i\omega{\upsilon_0^{(1)}}'+k r^2(\zeta_0+\zeta_1) h'-k^2r^2f\upsilon_0^{(1)} \right)\mathrm{d}r_*'\,+\\
    +\upsilon_0^{(2)}\int^{\infty}_{r_*} \frac{\upsilon_0^{(1)}}{W_{1,2}}   \left( 2i\omega{\upsilon_0^{(1)}}'+k r^2 (\zeta_0+\zeta_1)h'-k^2r^2f\upsilon_0^{(1)} \right)\mathrm{d}r_*'+\mathcal{O}(\omega k),\label{ysol}
\end{multline}
where $W_{1,2}$ is the Wronskian of $\upsilon_0^{(1)}$ and $\upsilon_0^{(2)}$. Now that we have the full solution, we can come back to our comment after Eqs.~(\ref{eq58}-\ref{eq59}) about the difficulties arising from the fact that $\gamma$ is not known in advance. The difficulties are resolved by the fact that all inhomogeneous equations contain a pure $\omega$-term (i.e., without multiplication by $k$), so whatever the scaling of $\omega$ with $k$, the leading terms with $\omega$ will be included at every order in $k$.


The solutions we have found consist of integrals which, in principle, have to be solved numerically. However, in the UV limit we can determine their primitive functions and thus obtain the asymptotic expansions of $\zeta$ and $\upsilon$. Since $e^{i\omega r_*}\to 1$ at the boundary, these are equivalent to the expansions of $Z$ and $Y$. Therefore we find that in this approximation:
\begin{equation}\label{Aconstants}
    U^{(0)}=\left(\frac{k^2}{3}-i\omega\right)\frac{k}{|k|}+\frac{2}{3}\hat{\mu}_0 k,\quad U^{(1)}=\frac{k}{|k|},\quad Y^{(0)}=\upsilon_0(0),\quad Y^{(1)}=c\upsilon_0'(0).
\end{equation}
Note that there is no boundary contribution from the integrals in \eqref{ysol}, since the terms linear in $r$ (i.e.~$r_*$) cancel out. Also, in both solutions we discard the terms of order $\omega k$ since we assume $\gamma\geq 1$, so $\omega k$ never dominates over $k^2$. If $\omega\sim k$, i.e. $\gamma=1$, the $k^2$ terms should also be neglected.

Finally, we give a brief comment on zero-temperature expansions. In principle, one might hope to repeat this procedure and find approximate low-energy solutions at $T=0$. However, this proves to be extremely challenging. The main difficulty arises when one has to cancel divergences in the IR, since $r\to\infty$ is a branch point. Namely, each term of the solution in the IR limit has a different (non-integer) scaling exponent, so that one cannot collect together the terms with the same exponent. A possible solution to this issue could be a regularization of the infinity; yet, it turns out to be very delicate and unreliable, and we decided not to pursue it. Instead, we can rely on our qualitative conclusion about the hydrodynamics at $T=0$ and estimate the diffusion constant numerically. We have found $D=0.14$ for $z=4$ and $D=0.41$ for $z=2.5$.

\subsection{Analytical Green's functions and instability}\label{analytical green's functions}

Plugging the relations \eqref{Aconstants} into Eqs.~\eqref{G+} and \eqref{G-}, we obtain approximate low-energy auxiliary Green's functions:
\begin{equation}
    G_+=-\frac{1-2Y^{(0)}\hat{Q}/(3\hat{M})|k|+C_0 k^2+\mathcal{O}(\omega k)}{i\omega+D_1 |k|-D_2 k^2+\mathcal{O}(k^3)},\label{GA+}
\end{equation}
\begin{equation}
    G_-=-\frac{Y^{(1)}}{Y^{(0)}}\frac{i\omega+D_1 |k|-(D_2+C_1)k^2+\mathcal{O}(\omega k)}{i\omega+D_1 |k|-D_2 k^2+\mathcal{O}(k^3)},\label{GA-}
\end{equation}
where
\begin{equation}
      C_{0,1}=\frac{2\hat{Q}\hat{\mu}_0}{9\hat{M}Y^{(0,1)}},\quad D_1=\frac{2}{3}\left(\frac{\hat{Q}}{\hat{M}} Y^{(0)}-\hat{\mu}_0\right),\quad D_2=\frac{1}{3}.
\end{equation}
We immediately notice that
\begin{equation}
    \lim_{\omega, k\to 0}G_-=-\frac{Y^{(1)}}{Y^{(0)}},
\end{equation}
implying that $G_-$ does not posses low-energy singularities (unless specifically $Y^{(0)}=0$). This observation agrees with the numerical results, and explains the somewhat featureless profile of the conductivity (remember that the conductivity is determined solely by $G_-$, according to Eq.~\eqref{3gyy}). Therefore, we are not much interested in studying $G_-$, and we focus on $G_+$ instead. 
The properties of the correlation functions (\ref{3gytyt}-\ref{3gxyy}) are then deduced straightforwardly (although one should be careful that multiplying an expression by $k^2$, $\omega^2$ or $\omega k$ changes its series expansion).

\begin{figure}[t]
    \begin{minipage}{0.5\textwidth}
        \centering
        \includegraphics[height=0.28\textheight]{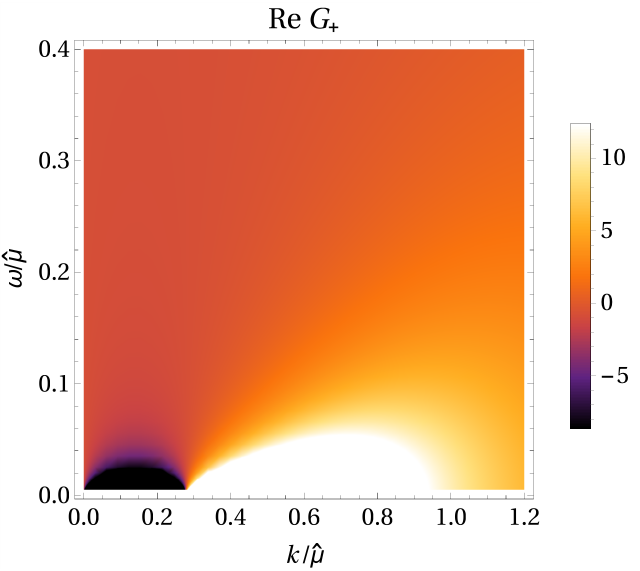}
    \end{minipage}
    \begin{minipage}{0.5\textwidth}
        \centering
        \includegraphics[height=0.28\textheight]{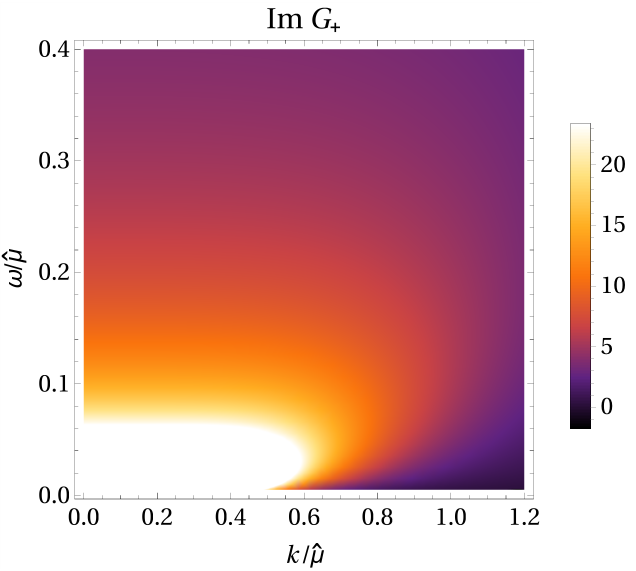}
    \end{minipage}\vspace{5pt}\\
    \begin{minipage}{0.5\textwidth}
        \centering
        \includegraphics[height=0.28\textheight]{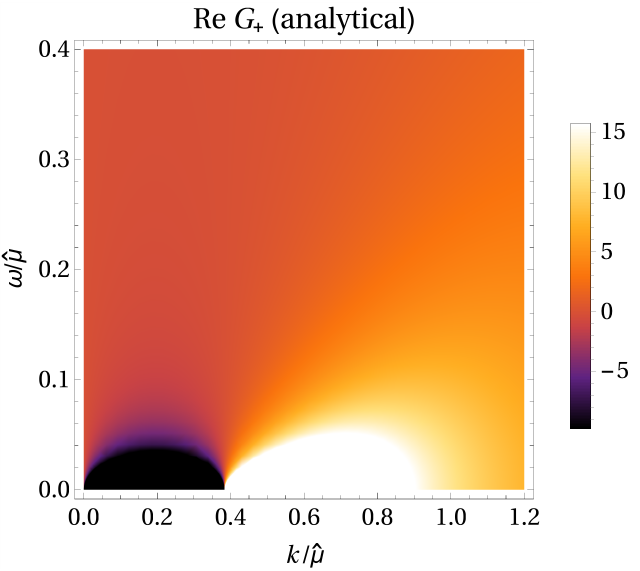}
    \end{minipage}
    \begin{minipage}{0.5\textwidth}
        \centering
        \includegraphics[height=0.28\textheight]{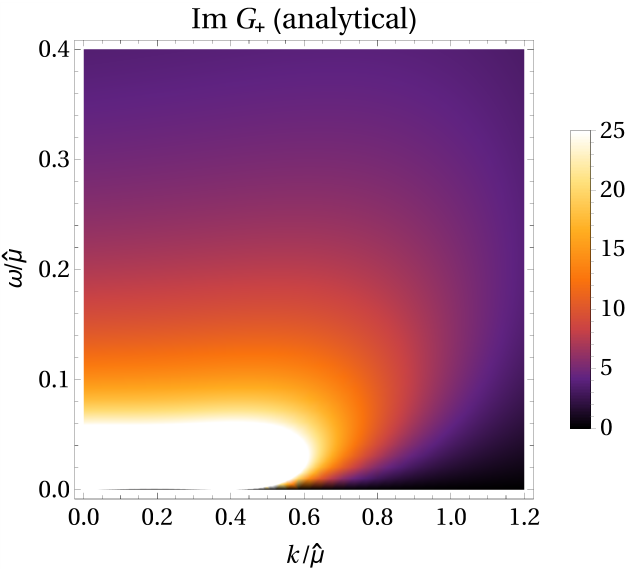}
    \end{minipage}
    \caption{Density plots of the real part (left) and the negative imaginary part (right) of $G_{ytyt}$ for $z=2.5$ and $T/T_c=0.95$. The figure shows a comparison between the numerical (top) and analytical results (bottom). The results agree pretty well.}
    \label{NvsA095}
\end{figure}
\begin{figure}[t]
    \begin{minipage}{0.5\textwidth}
        \centering
        \includegraphics[height=0.28\textheight]{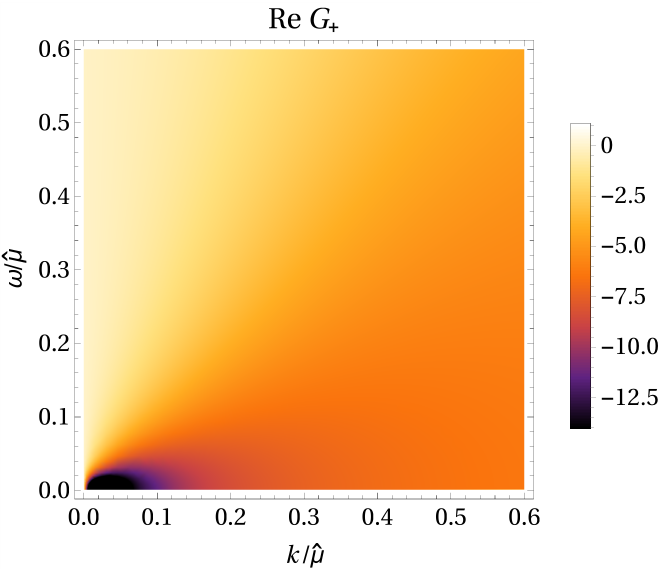}
    \end{minipage}
    \begin{minipage}{0.5\textwidth}
        \centering
        \includegraphics[height=0.28\textheight]{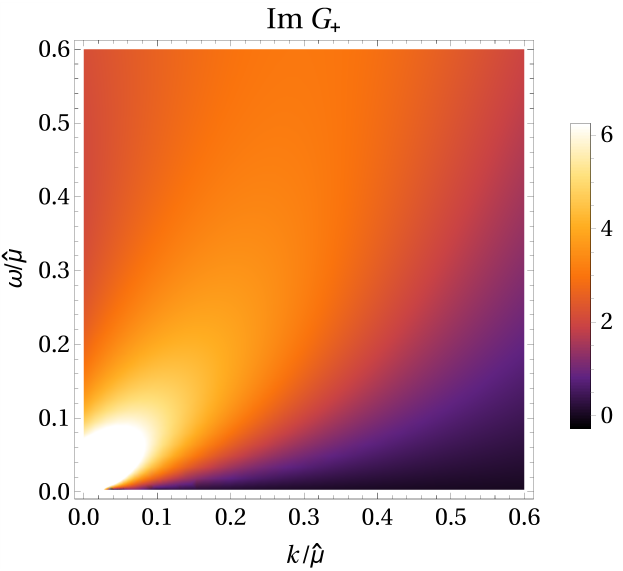}
    \end{minipage}\vspace{5pt}\\
    \begin{minipage}{0.5\textwidth}
        \centering
        \includegraphics[height=0.28\textheight]{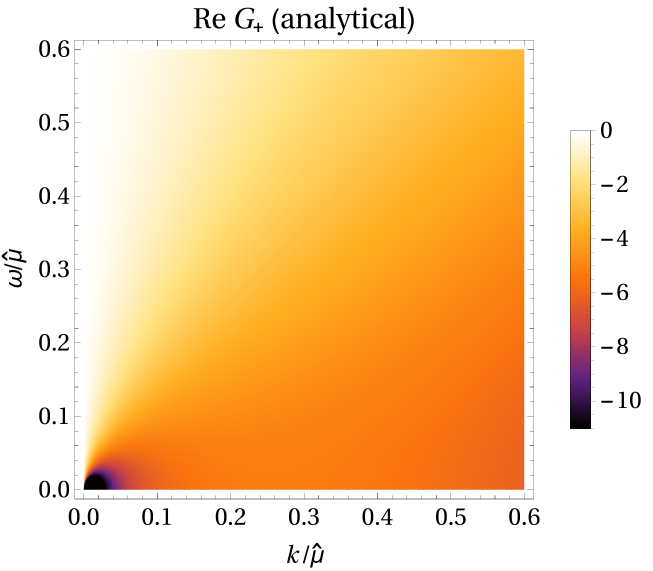}
    \end{minipage}
    \begin{minipage}{0.5\textwidth}
        \centering
        \includegraphics[height=0.28\textheight]{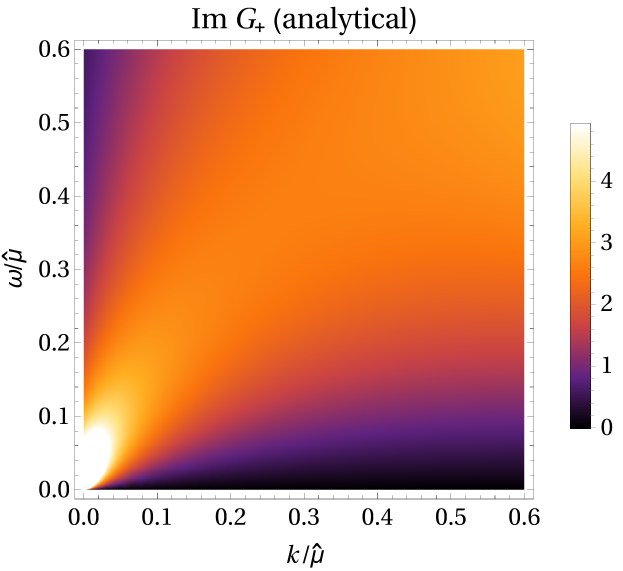}
    \end{minipage}
    \caption{Density plots of the real part (left) and the negative imaginary part (right) of $G_{ytyt}$ for $z=2.5$ and $T/T_c=0.65$. The figure shows a comparison between the numerical (top) and analytical results (bottom). The agreement is decent, although not as good as for $T/T_c=0.95$.}
    \label{NvsA065}
\end{figure}

\begin{table}[t]
    \centering
    \begin{tabular}{|c|r|r|r|}
     \hline\multicolumn{4}{|c|}{$z=2.5$}\\
     \hline $T/T_c$ & 0.35 & 0.65 & 0.95\\
     \hline $Y^{(0)}$ & 11183 & 7.12 & 1.38 \\ 
     \hline $Y^{(1)}$ & $-725815$ & $-16.18$ & $-0.46$\\
     \hline $D_1$ & 153.77 & 1.64 & 0.09\\\hline
\end{tabular}\hspace{10pt}
\begin{tabular}{|c|r|r|r|r|}
     \hline \multicolumn{5}{|c|}{$z=4$}\\
     \hline $T/T_c$ & 0.23 & 0.35 & 0.65 & 0.95\\
     \hline $Y^{(0)}$ & 5281 & 47.54 & 5.30 & 1.36 \\ 
     \hline $Y^{(1)}$ & $-88000$ & $-204.34$ & $-8.51$ & $-0.41$\\
     \hline $D_1$ & 251.42 & 7.48 & 1.30 & 0.07\\\hline
\end{tabular}\\
\vspace{10pt}
\begin{tabular}{|c|r|r|r|r|r|}
     \hline \multicolumn{6}{|c|}{$z=20$}\\
     \hline $T/T_c$ & 0.05 & 0.23 & 0.35 & 0.65 & 0.95\\
     \hline $Y^{(0)}$ & $-6.02\cdot10^{-8}$ & 22.65 & 11.41 & 3.65 & 1.32 \\ 
     \hline $Y^{(1)}$ & $1.20\cdot10^{-7}$& $-33.40$ & $-14.82$ & $-3.23$ & $-0.33$\\
     \hline $D_1$ & $-1.59$ & 6.77 & 3.26 & 0.75 & 0.04\\\hline
\end{tabular}
    \caption{The UV expansion coefficients of the gauge field $Y$ and the linear term coefficient from the dispersion relation obtained in the semianalytical approximation at finite temperatures.}
    \label{tab:d1}
\end{table}

We find that at the highest temperature $T/T_c=0.95$ the approximation \eqref{GA+} is in excellent, even quantitative agreement with the numerical results (Figure \ref{NvsA095}). At $T/T_c=0.65$ (Figure \ref{NvsA065}), the agreement is not as good, and as the temperature decreases further the analytical approximation becomes less and less satisfying. In general, the analytical and numerical results at low temperatures agree only for very small frequencies and momenta. This might have something to do with the fact that, as the temperature decreases, $Y^{(0)}$ and $Y^{(1)}$ drastically increase (Table~\ref{tab:d1}). 

But precisely the fact that $Y^{(0)}$ and $Y^{(1)}$ become very large can be exploited to learn something about the low-temperature correlation functions. In that regard, let us assume that there is a momentum scale $k_0$ ($k_0\ll 1$), such that $|k| Y^{(0,1)}\to\infty$ for $|k|\gg k_0$ and $|k| Y^{(0,1)}\to0$ for $|k|\ll k_0$ as $Y^{(0,1)}\to\infty$. Then:
\begin{equation}\label{GAlimit}
    G_+=\begin{cases}
            \frac{Y^{(1)}}{Y^{(0)}},\quad |k|\gg k_0\\
            \frac{1}{U^{(0)}},\quad |k|\ll k_0
        \end{cases}.
\end{equation}
From Eq.~\eqref{Aconstants} we see that for small enough momenta $G_+$ indeed develops a singular structure, while otherwise it becomes real. This might explain the agreement between the numerical and analytical results in the narrow domain of lowest energies, as well as the cases where at low energies the spectral function vanishes (see e.g.~Figure~\ref{dplot4_035_065}), thus supporting our claim.


Before we proceed with the analysis of properties of the correlation functions, a few words about $Y$ and its dependence on temperature are in order. Eq.~\eqref{yeq0} can be understood as a zero-energy Schr\"odinger equation for the perturbation $\upsilon_0\sim Y$. The potential $V$ is positive everywhere and its height grows considerably as the temperature decreases. Consequently, the amplitude of $Y$ must decrease with temperature as the field propagates through the potential towards the black hole. Since we set it to unity at the horizon in both the numerical and the analytical calculations, the integration of Eq.~\eqref{yeq0} must give us the boundary value of $Y$ that is commensurately rescaled. This explains why at low temperatures $Y^{(0)}$ and $Y^{(1)}$ are large. In what is to follow, we will see how the magnitude of $Y^{(0)}$ also controls the analytic structure of $G_+$.


Now we reap the main fruits of our semianalytic expansion. From Eqs.~\eqref{GA+} and \eqref{GA-} the dispersion relation reads:
\begin{equation}\label{dispersion}
    \omega=i D_1 |k|-i D_2 k^2+\mathcal{O}(k^3).
\end{equation}
There is both a linear and a quadratic term: the latter corresponds to diffusion and the former can either be a drift, if negative, or something else, if positive. In fact, it turns out that $D_1>0$ in most cases, so when $k\to 0$ there is a pole in the upper half-plane! For general $k$, the position and nature of the pole depend on the ratio of the linear and the quadratic term in \eqref{dispersion}. In other words, they depend on the competition between the two. It means that for each $D_1$ there exists a critical momentum
\begin{equation}
    k_c\equiv\frac{|D_1|}{D_2}=3|D_1|,
\end{equation}
which splits the two regimes:\footnote{Here we again recognize the issue we encountered when we were expanding the equations of motion in small $\epsilon$ in Eqs.~(\ref{zetaeq0}-\ref{yeq2}) when we had to deal with the ambiguity of $\gamma$, assuming $\omega\sim |k|^{\gamma}$.}
\begin{equation}
    \omega\sim\left\{ \begin{array}{ll}
        k,\quad & |k|<k_c\\
        k^2,\quad & |k|>k_c
    \end{array} \right. .
\end{equation}
Thus, if $|k|>k_c$, diffusive transport dominates; otherwise, the pole is in the upper half-plane and we have an instability. However, taking into account the numerical values of $D_1$ (Table~\ref{tab:d1}), we conclude that the competition between $D_1$ and $D_2$ only exists at high temperatures. As the temperature is reduced, $D_1$ (governed by $Y^{(0)}$) and consequently $k_c$ become so large that for $|k|>k_c$ the low-energy approximation no longer makes sense. Therefore, at intermediate temperatures, the low-momentum poles are expected to occur only in the upper half-plane.\footnote{Of course, this does not mean that there are no poles in the lower half-plane at all, only that they do not exist at small momenta.} At still lower temperatures, our approximation does not work very well, as we have seen, so we cannot analytically reproduce the re-entrance to the hydrodynamic regime seen in the numerics (Figures~\ref{linplot4_0_023} and \ref{dplot4_0_023}).

The presence of the linear term might explain why \cite{Kaplis:2013gux} finds an unexpected temperature dependence of the diffusion constant for certain parameter regimes: in \cite{Kaplis:2013gux} it is assumed that the dispersion relation describes conventional hydrodynamics, with purely quadratic diffusion. We find instead even more unexpected behavior, with a linear term that can give rise to an instability. On the other hand, \cite{Kaplis:2013gux} also studies the quasinormal modes, which we do not examine in this work.

\begin{figure}[t]
    \centering
    \includegraphics[width=\linewidth]{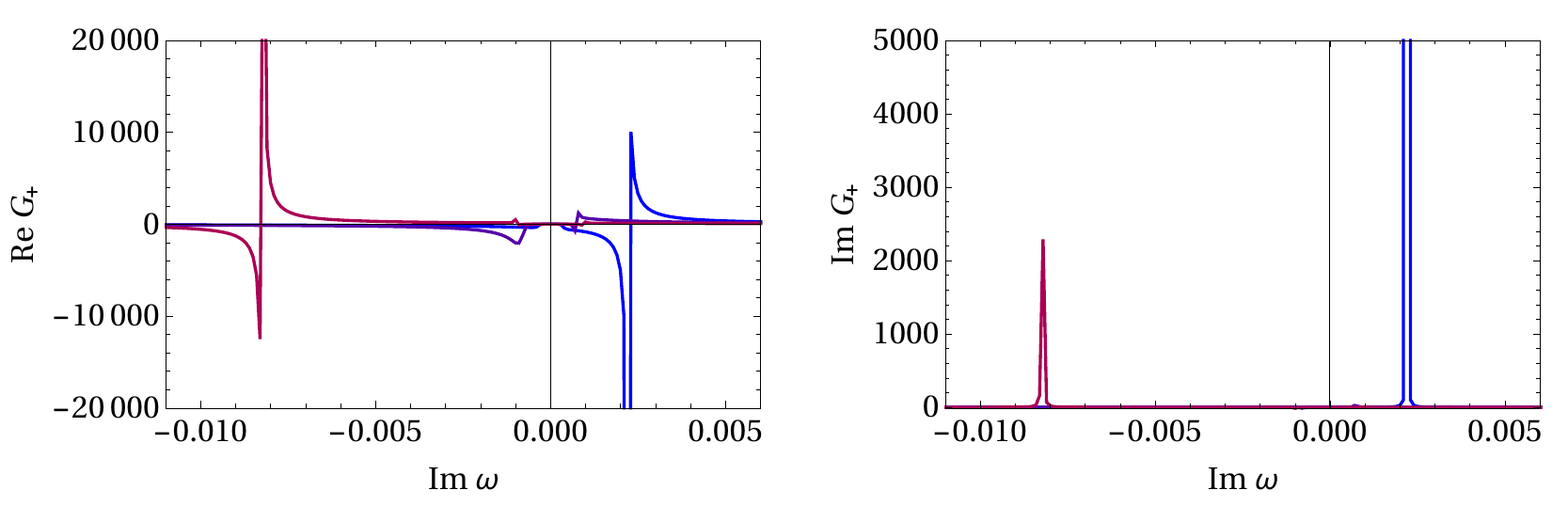}
    \includegraphics[width=\linewidth]{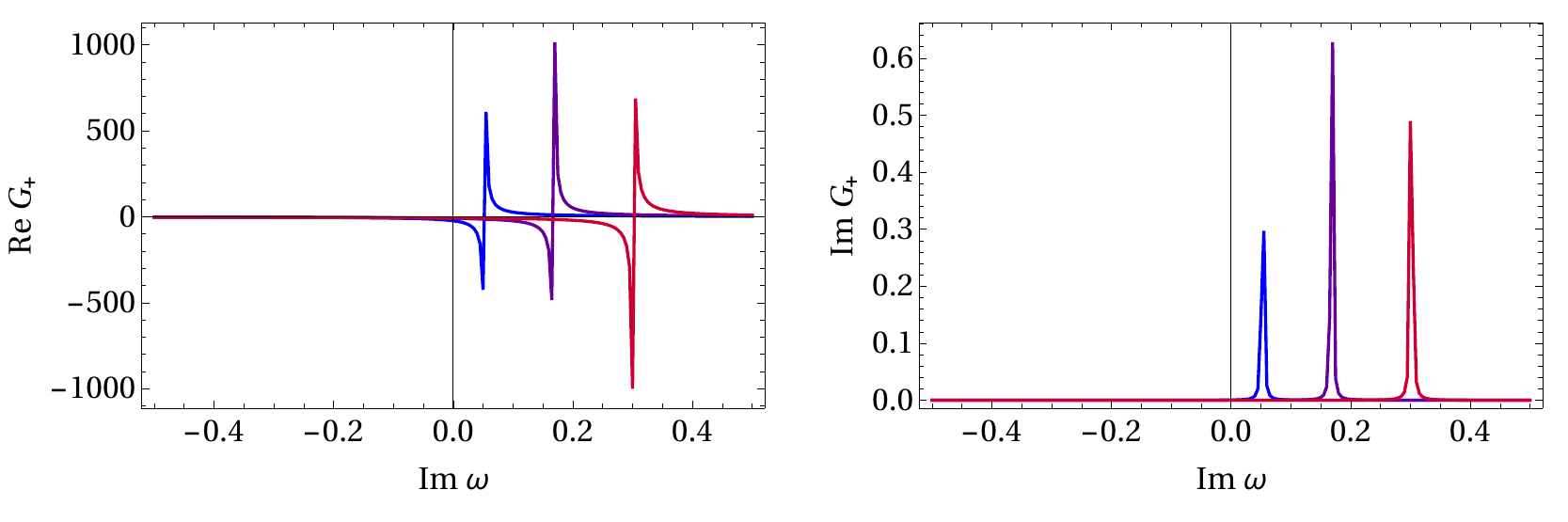}
    \includegraphics[width=\linewidth]{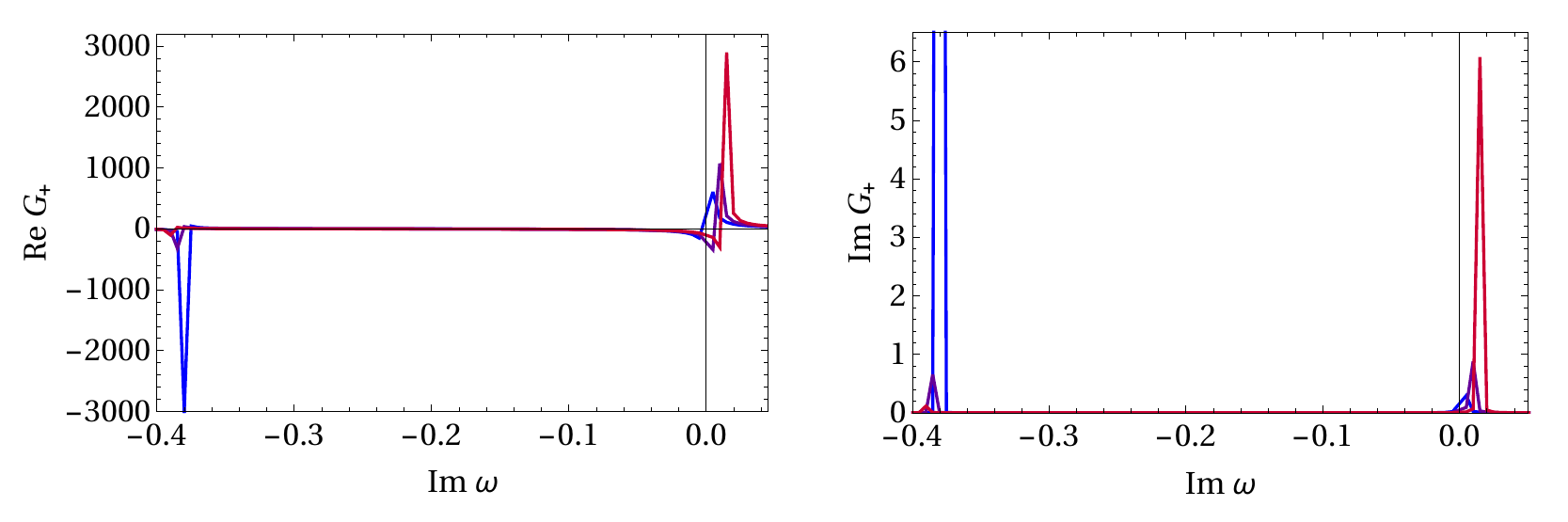}\vspace{-0.2cm}
    \caption{The pole structure of the real part (left) and the imaginary part (right) of $G_+$ for three temperature regimes: $T/T_c=0.95$, for $z=2.5$ and $k=0.1,0.2,0.3$ (from blue to red; top), $T/T_c=0.65$, for $z=4$ and $k=0.1,0.3,0.5$ (from blue to red; middle) and $T/T_c=0.23$, for $z=20$ and $k=0.01,0.03,0.05$ (from blue to red; bottom). The instability occurs at all temperatures. However, the intermediate-temperature low-energy poles arise exclusively in the upper half-plane!} 
    \label{figpoles}
\end{figure}

In order to check our analytical predictions, we numerically solve the fluctuation equations for a range of pure imaginary frequencies for a couple of backgrounds. In Figure \ref{figpoles} we show three representative examples for three temperature regimes, confirming that our conclusions are correct. What we could not predict analytically is that, as we mentioned above, diffusion poles do exist at low temperatures along with those in the upper half-plane (until at $T/T_c=0.05$ ($z=20$) only a diffusion pole remains). Although it could not be inferred from the preceding analysis, it is by no means surprising: according to what we argue in Section \ref{secresults} one actually does expect the hydrodynamics to reemerge at low temperatures, and these results support those observations. 

\begin{figure}[t]
    \centering
    \begin{minipage}[b]{0.5\textwidth}
    \centering
        \includegraphics[height=0.2\textheight]{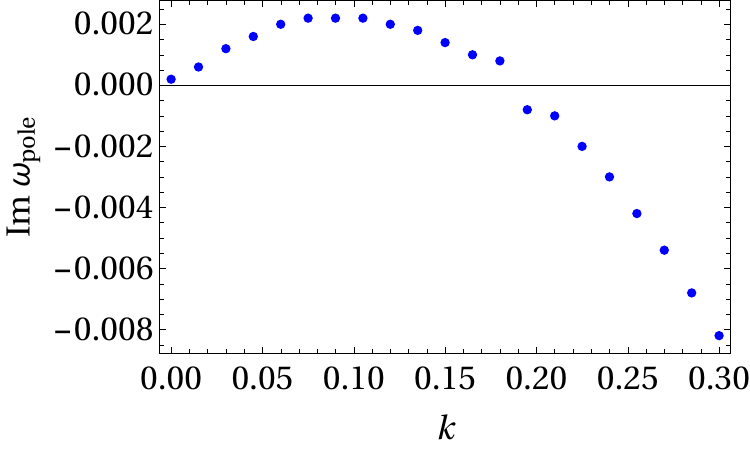}
    \end{minipage}%
    \begin{minipage}[b]{0.5\textwidth}
    \centering
        \includegraphics[height=0.2\textheight]{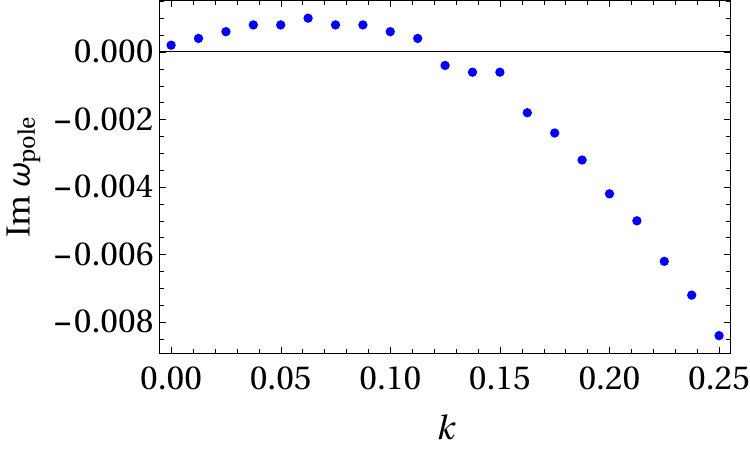}
    \end{minipage}\vspace{-0.2cm}
    \caption{Tracing of the pole of the auxiliary Green's function $G_+$ at $T/T_c=0.95$ for $z=2.5$ (left) and $z=4$ (right). We have found the pole positions by examining the maxima of $G_+$ at given $k$.}
    \label{pole tracing}
\end{figure}

Importantly, our instability lives in deep IR: the pole lies on the positive imaginary axis for arbitrarily small $k$. At high temperatures there is a critical value $k_c$ such that the pole moves to the negative imaginary axis for $|k|>k_c$, in accordance with our semianalytical prediction (Figure \ref{pole tracing}). At $T=0$ the pole never emerges in the upper half-plane, hence there is no instability.


Therefore, the electron cloud, while interpolating between two stable systems (the zero-temperature electron star and the RN black hole), corresponds to an unstable state in the boundary QFT. This is surprising: we know that the RN black hole itself has an instability in the presence of fermions, leading to the formation of a hairy black hole or (in the pure fluid limit) the electron star/cloud; it is unexpected that the cloud itself is not the true ground state. Although at this point we do not understand the meaning of this instability in field theory, we can gain at least some intuition in the bulk. The linear term in \eqref{dispersion} stems from the terms which couple $Z$ and $Y$ in the system (\ref{3eomszy1}-\ref{3eomszy2}).\footnote{Another way to understand this is to note that the coefficient $D_1$ depends mainly on $Y^{(0)}$, which is determined by the potential barrier $V$, which in turn stems from the presence of fermionic matter in the bulk and depends on temperature.} We have mentioned that for RN black hole the equations can be completely decoupled; therefore, it is the very presence of the cloud that gives rise to the linear term and eventually the instability.\footnote{This also means that the proof of stability from \cite{Horowitz:1999jd,Edalati:2010hk} does not apply to our system.} There must be a different finite-density solution where this is not the case; in the final section we will discuss what it might be.


\section{Discussion and conclusions}\label{secconc}

We have looked at the transverse channel of the linear response of the electron star, a simplified model of holographic Fermi surfaces. The significance of the electron star lies in the fact that it has provided insight into presumably stable fermionic holographic matter early on. The fluid limit taken in this model is somewhat pathological, and the field theory outcome -- an infinity of Fermi surfaces with exponentially small self-energies -- presents a highly unusual Fermi-liquid-like system of non-Landau type. Our goal was to understand better the physics of the system and the limits of the fluid (Thomas-Fermi) approximation.

It turns out that a big surprise lurks in this system. At high temperatures, near the transition point to the semilocal quantum liquid of the RN black hole, the response is expectedly hydrodynamic and close to that of RN. At low temperatures, the response is again dominated by the hydrodynamic diffusion; similar behavior is seen even at $T=0$, although here we expect the non-analytic contributions of the Lifshitz sector to modify the hydrodynamic pole.\footnote{We thank Blaise Gouteraux for discussions on this point.} But overall, this is not a big mystery either.

The big mystery is the presence of a pure imaginary pole in the upper frequency half-plane at intermediate temperatures. This signals an instability. Moreover, it is an IR instability (starting at arbitrarily small $k$), which implies that the electron cloud is not the true ground state, and hence likely not a viable holographic dual of a Fermi liquid at finite temperature. Understanding this instability is the main goal for future work in this direction. It might come (1) from the pathologies of the fluid limit -- once the non-classical tails and WKB corrections \cite{Medvedyeva:2013rpa} are taken into account, the system might become stable (2) simply from the existence of some other, thermodynamically preferred, solution of the (nonlinear) Einstein-Maxwell system (\ref{2eom1}-\ref{2eom3}).\footnote{One possibility is that the true solution is more compact while the electron star has a Jeans instability. This would nicely match with the IR nature of the instability. We thank Sean Hartnoll for this idea.} So far we have not been able to confirm either of the above scenarios; we also cannot discard other possibilities. In dual field theory this instability, present for a range of temperatures, could imply that the system corresponds to a quantum critical phase. One could expect that the instability should also manifest itself as a critical point or region in free energy. However, in order to see it one would need to add another operator which presumably condenses (acquires a VEV) at the transition point. Since we do not know what this operator is, we cannot perform this kind of analysis. In other words, we do not know what to compare with the star/cloud free energy. A possible alternative to thermodynamic analysis would be the study of quantum information measures such as those in \cite{DiNunno:2021eyf}.

We note in passing that the internal RN region of the electron cloud has been shown to be unstable to scalar hair formation in \cite{Carballo:2024hem}. It is not clear if our instability is related to this, but it does hint toward the conclusion that the configuration RN-cloud-RN is inherently fragile.


An encouraging aspect is the fact that low-$\omega$, low-$k$ expansion can be performed analytically and confirms all the key findings of the numerics, including the instability. This, together with the fact that we have reproduced the known ES-RN phase diagram and the known RN response from \cite{Edalati:2010hk,Davison:2011uk,Davison:2013bxa}, makes us confident in our findings. Achieving a full understanding of the mysterious instability is now an important task for future work. However, it also prompts us to look for a substantially different approach toward holographic (non-)Fermi liquids.

\acknowledgments

We are grateful to Blaise Gouteraux, Sa\v{s}o Grozdanov, Hareram Swain, Juan Pedraza, Sean Hartnoll and Ulf Gran for stimulating discussions. The work on Sections \ref{secintro}, \ref{secback} and \ref{secconc} (backgrounds, supervision and interpretation of results) was supported by Russian Science Foundation Grant No. 24-72-10061 [https://rscf.ru/project/24-72-10061/] and performed at Steklov Mathematical Institute of Russian Academy of Sciences (Mihailo \v{C}ubrovi\'c). The rest of the paper was funded by the Institute of Physics Belgrade, through the grant by the Ministry of Science, Technological Development, and Innovation of the Republic of Serbia.

\appendix

\section{Parity-invariance}\label{parity}

Let us define: $k h^x_{\phantom{x}y}(\omega,k;r)\equiv|k|\tilde{h}^x_{\phantom{x}y}(\omega,k;r)$, for $k\neq0$. We write explicit dependence on $\omega$ and $k$ in order to distinguish solutions corresponding to different sign of momentum. Then, if $k<0$, Eqs.~(\ref{eoms1}-\ref{3constraint}) give same solutions for $h^y_{\phantom{y}t}(\omega,k;r)$, $\tilde{h}^x_{\phantom{x}y}(\omega,k;r)$ and $a_y(\omega,k;r)$ as if $k$ were positive:
\begin{equation}
    h^y_{\phantom{y}t}(\omega,k;r)=h^y_{\phantom{y}t}(\omega,|k|;r),\quad \tilde{h}^x_{\phantom{x}y}(\omega,k;r)=h^x_{\phantom{x}y}(\omega,|k|;r),\quad a_y(\omega,k;r)=a_y(\omega,|k|;r),
\end{equation}
which means that $h^x_{\phantom{x}y}(\omega,k;r)=-\tilde{h}^x_{\phantom{x}y}(\omega,k;r)=-h^x_{\phantom{x}y}(\omega,|k|;r)$ when $k<0$. Same logic applies to $X$, $Z$ and $U$. Then the asymptotic master fields \eqref{3masterfields} and, consequently, the auxiliary Green's functions \eqref{3gpm}, as well as the boundary action (\ref{3sbnd}-\ref{3sbnd2}) are invariant under the exchange $k\to-k$. Alternatively, we could have defined $k h^y_{\phantom{y}t}(\omega,k;r)=|k|\tilde{h}^y_{\phantom{y}t}(\omega,k;r)$ and $k a_y(\omega,k;r)=|k|\tilde{a}_y(\omega,k;r)$, in which case $Y$ and $\Phi_{\pm}$ would change their sign. In both cases the theory remains parity-invariant.

\section{UV asymptotics}\label{UVap}

The raw, gauge-dependent expansions read:
\begin{eqnarray}
    h^x_{\phantom{x}y}(r\to0)&=&h^{(0)}_{x y}+\frac{\omega}{2c^2}(\omega h^{(0)}_{x y}+k h^{(0)}_{y t})r^2+h^{(3)}_{x y}r^3+\ldots\label{UVxy}\\
    h^y_{\phantom{y}t}(r\to0)&=&h^{(0)}_{y t}-\frac{k}{2}(\omega h^{(0)}_{x y}+k h^{(0)}_{y t})r^2+\left(\frac{2}{3}c\hat{Q}a^{(0)}_y-\frac{c^2k}{\omega}h^{(3)}_{x y} \right)r^3+\ldots\label{UVyt}\\
    a_y(r\to0)&=&a^{(0)}_y+a^{(1)}_y r+\frac{a^{(0)}_y}{2}\left(k^2-\frac{\omega^2}{c^2}\right)r^2+\ldots,\label{UVay}
\end{eqnarray}
leading to the following expansions for the gauge-invariant fields:
\begin{eqnarray}
    X(r\to0)&=&X^{(0)}-\frac{X^{(0)}}{2}\left(k^2-\frac{\omega^2}{c^2}\right)r^2+X^{(3)}r^3+\ldots,\\
    Y(r\to0)&=&Y^{(0)}+Y^{(1)} r+\frac{Y^{(0)}}{2}\left(k^2-\frac{\omega^2}{c^2}\right)r^2+\ldots,
\end{eqnarray}
where 
\begin{equation}
    X^{(0)}=k h^{(0)}_{y t}+\omega h^{(0)}_{x y},\quad X^{(3)}=\frac{2}{3}c \hat{Q} k a^{(0)}_y+\frac{1}{\omega}(\omega^2-c^2k^2)h^{(3)}_{x y},\quad a_y^{(0, 1)}=Y^{(0, 1)}. 
\end{equation}
In a similar manner we obtain
\begin{equation}
    U(r\to0)=U^{(0)}+U^{(1)}r+\frac{U^{(0)}}{2}\left( k^2-\frac{\omega^2}{c^2} \right)r^2+\ldots,
\end{equation}
where
\begin{equation}
    U^{(0)}=-\frac{X^{(0)}}{c},\quad U^{(1)}=-\frac{2k c^2\hat{Q}Y^{(0)}-3c X^{(3)}}{(c^2k^2-\omega^2)}.
\end{equation}
All higher-order coefficients are determined straightforwardly. We have obtained the expansions up to eighth order and we keep them in \verb|Mathematica| notebooks.

We use these expansions in order to determine the boundary action. The renormalized, on-shell boundary action (obtained by using the prescription of \cite{Kaminski:2009dh}) reads
\begin{multline}
    S^\mathrm{on-shell}_\mathrm{bdy}=\lim_{r\to 0}\frac{L^2}{\kappa^2}\int\frac{\mathrm{d}\omega}{2\pi}\int\frac{\mathrm{d}k}{2\pi}\int\mathrm{d}y \Big( a_y(-\omega, -k) \left(r^2 f a_y'(\omega, k)+h' h^y_{\phantom{y}t}(\omega, k) \right)+ \\
    +\frac{1}{2r^2}\left( h^x_{\phantom{x}y}(-\omega, -k) (r^2 f h^x_{\phantom{x}y}(\omega, k) )'-h^y_{\phantom{y}t}(-\omega, -k) {h^y_{\phantom{y}t}}'(\omega, k) \right)-\\
    -\frac{2}{r^3}\left( 1-\frac{c}{r\sqrt{f}} \right) \left( r^2 f h^x_{\phantom{x}y}(-\omega, -k)h^x_{\phantom{x}y}(\omega, k)-h^y_{\phantom{y}t}(-\omega, -k)h^y_{\phantom{y}t}(\omega, k) \right)-\\
    -\frac{c}{2r^2\sqrt{f}}(\omega h^x_{\phantom{x}y}(-\omega, -k)+k h^y_{\phantom{y}t}(-\omega, -k))(\omega h^x_{\phantom{x}y}(\omega, k)+k h^y_{\phantom{y}t}(\omega, k)) \Big).\label{3sbnd}
\end{multline}
Upon Taylor expanding \eqref{3sbnd} in small $r$ and applying the constraint \eqref{3constraint}, we get
\begin{multline}
    S^{\mathrm{on-shell}}_\mathrm{bdy}=\frac{L^2 c^2}{\kappa^2}\int\frac{\mathrm{d}\omega}{2\pi}\int\frac{\mathrm{d}k}{2\pi}\int\mathrm{d}y \Big( a^{(0)}_y(-\omega, -k) a^{(1)}_y(\omega, k)\,+\\
    +\frac{3}{2}h^{(0)}_{xy}(-\omega, -k) h^{(3)}_{xy}(\omega, k)+\frac{3k}{2\omega}h^{(0)}_{yt}(-\omega, -k) h^{(3)}_{xy}(\omega, k) \Big)+\textrm{contact terms}.\label{3sbnd2}
\end{multline}
We use the auxiliary Green's functions (\ref{G+}-\ref{G-}) to express the responses $h^{(3)}_{xy}$ and $a^{(1)}_y$ in terms of $h^{(0)}_{xy}$, $h^{(0)}_{yt}$ and $a^{(0)}_y$ and to obtain an action that is bilinear in sources. Finally, we take its functional derivatives in order to get the field-theory correlation functions.

\section{IR asymptotics}\label{secflucir}

The IR boundary conditions differ between the zero-temperature and finite-temperature star. At finite temperature the far IR is just the inner RN region, and we can employ the Frobenius method, following the standard procedure for an RN horizon. For any fluctuating quantity, denoted generically by $C(r)$, we assume solutions in form
\begin{equation}
    C(r)=(r_h-r)^{-i\nu}\sum_{n=0}^{\infty}c_n (r_h-r)^n.\label{irexp}
\end{equation}
The branch-point exponent $\nu$ is found by solving the indicial polynomial and reads $\nu=\omega/(4\pi c T)$. The expansion coefficients $c_n$ are then found by solving the equations order by order. The result is of course identical to the previous calculations in \cite{Edalati:2010hk,Davison:2011uk,Davison:2013bxa}, so there is no reason to repeat it here. We normalize the solutions setting $c_0=1$. We have determined the expansions up to sixth order.

The $T=0$ case is slightly more involved. We solve the gauge dependent equations (\ref{eoms1}-\ref{eoms3}) in the limit $r\to \infty$, taking into account that $z>1$. At leading order we obtain
\begin{equation}
    a_y=A_0 e^{i\omega\sqrt{g_{\infty}}r^z/z},\quad h^y_{\phantom{y}t}=B_0 r^{3-2z}e^{i\omega\sqrt{g_{\infty}}r^z/z},\quad h^x_{\phantom{x}y}=-B_0\frac{\omega}{k} r e^{i\omega\sqrt{g_{\infty}}r^z/z},
\end{equation}
where $A_0$ and $B_0$ are arbitrary complex constants. 
In the deep IR limit we are able to determine $X$ and $Y$ with power series corrections to the leading order assuming that the functions take the following form
\begin{equation}
    C(r)=S_1(r) e^{i\sqrt{g_{\infty}}\omega r^z/z+S_2(r)},\label{3lifshtzsol}
\end{equation}
where $S_1$ and $S_2$ are series expansions in $r$ with non-analytic powers. The powers and series coefficients are again determined order by order by solving a system of algebraic equations. However, several difficulties arise because different powers of $r$ compete depending on the value of $z$, and one is unable to order the corrections from lower to higher for general $z$. In addition, higher-order terms in the expansion (\ref{3lifshtzsol}) also require higher-order corrections to the background itself to be taken into account for the whole calculation to be consistent. 
Fortunately, these complications turn out to be irrelevant because it is much more convenient to solve the equations with $U$ and $Y$ (Eqs.~\ref{3eomsuy1}-\ref{3eomsuy2}). This set of equations is simpler than the original one; finding the IR behavior is now much easier, and we can include the background corrections from the beginning. We follow the logic of \cite{Hartnoll:2010gu} and find the corrections as a power series in $r^{\alpha}$. We obtain: 
\begin{equation}
    U(r\to\infty)=Y(r\to\infty)=\left(1+\frac{i\sqrt{g_{\infty}}\omega(3-2g_1-g_1^2)r^{z+2\alpha}}{8(z+2\alpha)}+\ldots\right)e^{i\sqrt{g_{\infty}}\omega\left(\frac{r^z}{z}+\frac{(1+g_1)r^{z+\alpha}}{2(z+\alpha)}\right)}.
\end{equation}



\bibliography{RefsFL.bib}

\providecommand{\href}[2]{#2}\begingroup\raggedright\begin{thebibliography}{10}

\bibitem{Vegh:2009}
H.~Liu, J.~McGreevy, and D.~Vegh, {\it {Non-Fermi liquids from holography}},  {\em Phys. Rev. D} {\bf 83} (2011) 065029, [\href{http://arxiv.org/abs/0903.2477}{{\tt arXiv:0903.2477}}].

\bibitem{Leiden:2009}
M.~\v{C}ubrovi\'c, J.~Zaanen, and K.~Schalm, {\it {String Theory, Quantum Phase Transitions and the Emergent Fermi-Liquid}},  {\em Science} {\bf 325} (2009) 439--444, [\href{http://arxiv.org/abs/0904.1993}{{\tt arXiv:0904.1993}}].

\bibitem{Faulkner:2010zz}
T.~Faulkner, N.~Iqbal, H.~Liu, J.~McGreevy, and D.~Vegh, {\it {Strange metal transport realized by gauge/gravity duality}},  {\em Science} {\bf 329} (2010) 1043--1047.

\bibitem{Liu:2011}
N.~Iqbal, H.~Liu, and M.~Mezei, {\it {Semi-local quantum liquids}},  {\em JHEP} {\bf 04} (2012) 086, [\href{http://arxiv.org/abs/1105.4621}{{\tt arXiv:1105.4621}}].

\bibitem{Gubser:2009qt}
S.~S. Gubser and F.~D. Rocha, {\it {Peculiar properties of a charged dilatonic black hole in $AdS_{5}$}},  {\em Phys. Rev. D} {\bf 81} (2010) 046001, [\href{http://arxiv.org/abs/0911.2898}{{\tt arXiv:0911.2898}}].

\bibitem{Huijse:2011hp}
L.~Huijse and S.~Sachdev, {\it {Fermi surfaces and gauge-gravity duality}},  {\em Phys. Rev. D} {\bf 84} (2011) 026001, [\href{http://arxiv.org/abs/1104.5022}{{\tt arXiv:1104.5022}}].

\bibitem{Smit:2021dwh}
S.~Smit et~al., {\it {Momentum-dependent scaling exponents of nodal self-energies measured in strange metal cuprates and modelled using semi-holography}},  {\em Nature Commun.} {\bf 15} (2024), no.~1 4581, [\href{http://arxiv.org/abs/2112.06576}{{\tt arXiv:2112.06576}}].

\bibitem{Hercek:2022wyu}
F.~Her\v{c}ek, V.~Gecin, and M.~\v{C}ubrovi\'c, {\it {Photoemission ''experiments'' on holographic lattices}},  {\em SciPost Phys. Core} {\bf 6} (2023) 027, [\href{http://arxiv.org/abs/2208.05920}{{\tt arXiv:2208.05920}}].

\bibitem{Balm:2022bju}
F.~Balm et~al., {\it {T-linear resistivity, optical conductivity, and Planckian transport for a holographic local quantum critical metal in a periodic potential}},  {\em Phys. Rev. B} {\bf 108} (2023), no.~12 125145, [\href{http://arxiv.org/abs/2211.05492}{{\tt arXiv:2211.05492}}].

\bibitem{Charmousis:2010zz}
C.~Charmousis, B.~Gouteraux, B.~S. Kim, E.~Kiritsis, and R.~Meyer, {\it {Effective Holographic Theories for low-temperature condensed matter systems}},  {\em JHEP} {\bf 11} (2010) 151, [\href{http://arxiv.org/abs/1005.4690}{{\tt arXiv:1005.4690}}].

\bibitem{Gouteraux:2011ce}
B.~Gouteraux and E.~Kiritsis, {\it {Generalized Holographic Quantum Criticality at Finite Density}},  {\em JHEP} {\bf 12} (2011) 036, [\href{http://arxiv.org/abs/1107.2116}{{\tt arXiv:1107.2116}}].

\bibitem{Gouteraux:2012yr}
B.~Gouteraux and E.~Kiritsis, {\it {Quantum critical lines in holographic phases with (un)broken symmetry}},  {\em JHEP} {\bf 04} (2013) 053, [\href{http://arxiv.org/abs/1212.2625}{{\tt arXiv:1212.2625}}].

\bibitem{Zaanen:2021llz}
J.~Zaanen, {\it {Lectures on quantum supreme matter}},  \href{http://arxiv.org/abs/2110.00961}{{\tt arXiv:2110.00961}}.

\bibitem{Hartnoll:2010gu}
S.~A. Hartnoll and A.~Tavanfar, {\it {Electron stars for holographic metallic criticality}},  {\em Phys. Rev. D} {\bf 83} (2011) 046003, [\href{http://arxiv.org/abs/1008.2828}{{\tt arXiv:1008.2828}}].

\bibitem{Kaplis:2013gux}
N.~Kaplis, {\em {Holographic quantum liquids}}.
\newblock PhD thesis, Oxford U., 2013.

\bibitem{Gran:2018jnt}
U.~Gran, M.~Torns\"o, and T.~Zingg, {\it {Holographic Response of Electron Clouds}},  {\em JHEP} {\bf 03} (2019) 019, [\href{http://arxiv.org/abs/1810.11416}{{\tt arXiv:1810.11416}}].

\bibitem{Gran:2017jht}
U.~Gran, M.~Torns\"o, and T.~Zingg, {\it {Holographic Plasmons}},  {\em JHEP} {\bf 11} (2018) 176, [\href{http://arxiv.org/abs/1712.05672}{{\tt arXiv:1712.05672}}].

\bibitem{Gran:2018vdn}
U.~Gran, M.~Torns\"o, and T.~Zingg, {\it {Exotic Holographic Dispersion}},  {\em JHEP} {\bf 02} (2019) 032, [\href{http://arxiv.org/abs/1808.05867}{{\tt arXiv:1808.05867}}].

\bibitem{Edalati:2010hk}
M.~Edalati, J.~I. Jottar, and R.~G. Leigh, {\it {Shear Modes, Criticality and Extremal Black Holes}},  {\em JHEP} {\bf 04} (2010) 075, [\href{http://arxiv.org/abs/1001.0779}{{\tt arXiv:1001.0779}}].

\bibitem{Davison:2011uk}
R.~A. Davison and N.~K. Kaplis, {\it {Bosonic excitations of the $AdS_4$ Reissner-Nordstrom black hole}},  {\em JHEP} {\bf 12} (2011) 037, [\href{http://arxiv.org/abs/1111.0660}{{\tt arXiv:1111.0660}}].

\bibitem{Davison:2013bxa}
R.~A. Davison and A.~Parnachev, {\it {Hydrodynamics of cold holographic matter}},  {\em JHEP} {\bf 06} (2013) 100, [\href{http://arxiv.org/abs/1303.6334}{{\tt arXiv:1303.6334}}].

\bibitem{Davison:2013uha}
R.~A. Davison, M.~Goykhman, and A.~Parnachev, {\it {AdS/CFT and Landau Fermi liquids}},  {\em JHEP} {\bf 07} (2014) 109, [\href{http://arxiv.org/abs/1312.0463}{{\tt arXiv:1312.0463}}].

\bibitem{Sybesma:2015oha}
W.~Sybesma and S.~Vandoren, {\it {Lifshitz quasinormal modes and relaxation from holography}},  {\em JHEP} {\bf 05} (2015) 021, [\href{http://arxiv.org/abs/1503.07457}{{\tt arXiv:1503.07457}}].

\bibitem{Gursoy:2016tgf}
U.~G\"ursoy, A.~Jansen, W.~Sybesma, and S.~Vandoren, {\it {Holographic Equilibration of Nonrelativistic Plasmas}},  {\em Phys. Rev. Lett.} {\bf 117} (2016), no.~5 051601, [\href{http://arxiv.org/abs/1602.01375}{{\tt arXiv:1602.01375}}].

\bibitem{Hartnoll:2010ik}
S.~A. Hartnoll and P.~Petrov, {\it {Electron star birth: A continuous phase transition at nonzero density}},  {\em Phys. Rev. Lett.} {\bf 106} (2011) 121601, [\href{http://arxiv.org/abs/1011.6469}{{\tt arXiv:1011.6469}}].

\bibitem{GiangrecoMPuletti:2015grq}
V.~Giangreco M.~Puletti, S.~Nowling, L.~Thorlacius, and T.~Zingg, {\it {Magnetic oscillations in a holographic liquid}},  {\em Phys. Rev. D} {\bf 91} (2015), no.~8 086008, [\href{http://arxiv.org/abs/1501.06459}{{\tt arXiv:1501.06459}}].

\bibitem{Hartnoll:2009ns}
S.~A. Hartnoll, J.~Polchinski, E.~Silverstein, and D.~Tong, {\it {Towards strange metallic holography}},  {\em JHEP} {\bf 04} (2010) 120, [\href{http://arxiv.org/abs/0912.1061}{{\tt arXiv:0912.1061}}].

\bibitem{Leiden:2011}
M.~\v{C}ubrovi\'c, Y.~Liu, K.~Schalm, Y.-W. Sun, and J.~Zaanen, {\it {Spectral probes of the holographic Fermi groundstate: dialing between the electron star and AdS Dirac hair}},  {\em Phys. Rev. D} {\bf 84} (2011) 086002, [\href{http://arxiv.org/abs/1106.1798}{{\tt arXiv:1106.1798}}].

\bibitem{Emparan:1999pm}
R.~Emparan, C.~V. Johnson, and R.~C. Myers, {\it {Surface terms as counterterms in the AdS / CFT correspondence}},  {\em Phys. Rev. D} {\bf 60} (1999) 104001, [\href{http://arxiv.org/abs/hep-th/9903238}{{\tt hep-th/9903238}}].

\bibitem{Faulkner:2009wj}
T.~Faulkner, H.~Liu, J.~McGreevy, and D.~Vegh, {\it {Emergent quantum criticality, Fermi surfaces, and AdS(2)}},  {\em Phys. Rev. D} {\bf 83} (2011) 125002, [\href{http://arxiv.org/abs/0907.2694}{{\tt arXiv:0907.2694}}].

\bibitem{Jansen:2019wag}
A.~Jansen, A.~Rostworowski, and M.~Rutkowski, {\it {Master equations and stability of Einstein-Maxwell-scalar black holes}},  {\em JHEP} {\bf 12} (2019) 036, [\href{http://arxiv.org/abs/1909.04049}{{\tt arXiv:1909.04049}}].

\bibitem{Kaminski:2009dh}
M.~Kaminski, K.~Landsteiner, J.~Mas, J.~P. Shock, and J.~Tarrio, {\it {Holographic Operator Mixing and Quasinormal Modes on the Brane}},  {\em JHEP} {\bf 02} (2010) 021, [\href{http://arxiv.org/abs/0911.3610}{{\tt arXiv:0911.3610}}].

\bibitem{Policastro:2002se}
G.~Policastro, D.~T. Son, and A.~O. Starinets, {\it {From AdS / CFT correspondence to hydrodynamics}},  {\em JHEP} {\bf 09} (2002) 043, [\href{http://arxiv.org/abs/hep-th/0205052}{{\tt hep-th/0205052}}].

\bibitem{Policastro:2002tn}
G.~Policastro, D.~T. Son, and A.~O. Starinets, {\it {From AdS / CFT correspondence to hydrodynamics. 2. Sound waves}},  {\em JHEP} {\bf 12} (2002) 054, [\href{http://arxiv.org/abs/hep-th/0210220}{{\tt hep-th/0210220}}].

\bibitem{Kovtun:2005ev}
P.~K. Kovtun and A.~O. Starinets, {\it {Quasinormal modes and holography}},  {\em Phys. Rev. D} {\bf 72} (2005) 086009, [\href{http://arxiv.org/abs/hep-th/0506184}{{\tt hep-th/0506184}}].

\bibitem{Starinets:2008fb}
A.~O. Starinets, {\it {Quasinormal spectrum and the black hole membrane paradigm}},  {\em Phys. Lett. B} {\bf 670} (2009) 442--445, [\href{http://arxiv.org/abs/0806.3797}{{\tt arXiv:0806.3797}}].

\bibitem{Horowitz:1999jd}
G.~T. Horowitz and V.~E. Hubeny, {\it {Quasinormal modes of AdS black holes and the approach to thermal equilibrium}},  {\em Phys. Rev. D} {\bf 62} (2000) 024027, [\href{http://arxiv.org/abs/hep-th/9909056}{{\tt hep-th/9909056}}].

\bibitem{Medvedyeva:2013rpa}
M.~V. Medvedyeva, E.~Gubankova, M.~\v{C}ubrovi\'c, K.~Schalm, and J.~Zaanen, {\it {Quantum corrected phase diagram of holographic fermions}},  {\em JHEP} {\bf 12} (2013) 025, [\href{http://arxiv.org/abs/1302.5149}{{\tt arXiv:1302.5149}}].

\bibitem{DiNunno:2021eyf}
B.~S. DiNunno, N.~Jokela, J.~F. Pedraza, and A.~P\"onni, {\it {Quantum information probes of charge fractionalization in large-$N$ gauge theories}},  {\em JHEP} {\bf 05} (2021) 149, [\href{http://arxiv.org/abs/2101.11636}{{\tt arXiv:2101.11636}}].

\bibitem{Carballo:2024hem}
J.~Carballo, A.~K. Patra, and J.~F. Pedraza, {\it {Diving inside holographic metals}},  \href{http://arxiv.org/abs/2408.07748}{{\tt arXiv:2408.07748}}.

\end{thebibliography}\endgroup
\end{document}